\documentclass[a4paper,11pt]{article}

\usepackage{preamble}

\title{Neutrino oscillation data and a pseudo-Dirac heavy neutral lepton}

\author{Jan Hajer\orcmail{0000-0001-8083-9102}{jan.hajer@tecnico.ulisboa.pt}}

\affiliation{Departamento de Física, Instituto Superior Técnico (IST), Universidade de Lisboa, 1049-001 Lisboa, Portugal}

\usepackage{cite}

\begin{document}

\maketitle

\begin{abstract}
Symmetry-protected seesaw models can accommodate light-neutrino oscillation data while keeping \HNLs within collider reach.
In these models, the smallness of the light-neutrino masses is protected by an approximate \LN-like symmetry that is broken only by small parameters.
We study the minimal scenario in which the new states form one pseudo-Dirac \HNL pair.
The exact \LN-conserving Dirac limit is diagonalised without expanding in the active-sterile mixing, and the small \LN-violating entries are then included perturbatively.
This yields a symmetry-protected flavour reconstruction of the active-heavy interaction matrix.
The rank-two light-neutrino mass matrix fixes the normalised active-flavour direction, while the remaining high-energy information is a single complex light-heavy amplitude whose phase defines a $\CP$-odd light-heavy invariant.
For the normalised leading active-heavy interaction weights, this amplitude and the heavy-sector rotation cancel, leaving an ellipse in the flavour simplex determined by light-neutrino oscillation data and the Majorana phase.
We also identify how the linear \LN-violating terms enter coherent heavy-neutrino oscillations and the \NDBlong effective mass.
\end{abstract}

\clearpage

\tableofcontents

\listoftables

\listoffigures

\clearpage

\section{Introduction} \label{sec:introduction}

Neutrino oscillations have established that lepton flavours mix and, therefore, that at least two neutrinos are massive \cite{Super-Kamiokande:1998kpq, SNO:2002tuh, KamLAND:2002uet}.
Measurements of solar, atmospheric, reactor, and accelerator neutrinos have determined two independent mass-squared differences, large mixing angles, and a \CP-violating phase \cite{K2K:2004iot,MINOS:2006foh,T2K:2015sqm}.
The \SM, however, predicts massless neutrinos, since it contains no renormalisable operator that can generate neutrino masses.
The origin of neutrino masses is therefore one of the clearest indications that the particle content or symmetries of the \SM must be extended.

The most direct extension adds right-handed neutrinos while preserving \LN.
After \EWSB, their Yukawa interactions with the lepton doublets and the Higgs field generate Dirac neutrino masses, in analogy with the charged fermions.
Reproducing the observed neutrino masses in this way requires Yukawa couplings many orders of magnitude below those of the other \SM fermions.
An alternative construction invokes the dimension-five Weinberg operator, which violates \LN \cite{Weinberg:1979sa}.
After \EWSB, this operator generates Majorana masses suppressed by a heavy \NP scale.
The seesaw mechanisms provide \UV completions of this operator \cite{Minkowski:1977sc, Gell-Mann:1979vob, Yanagida:1979as, Glashow:1979nm, Mohapatra:1979ia,Schechter:1980gr}.

In conventional \TISs, the smallness of the light-neutrino masses is ensured by the heaviness of the new states or the smallness of their interaction strength.
This conclusion can be relaxed if the light-neutrino masses arise from cancellations between different entries of the neutrino mass matrix \cite{Kersten:2007vk}.
In that case, sizeable Yukawa couplings and active-sterile mixings can coexist with light-neutrino masses, making \HNL states experimentally accessible while remaining consistent with oscillation data.

Such cancellations can be protected by an approximate \LN symmetry.
In the exact \LN-conserving limit, the light neutrinos remain massless, but sizeable active-sterile mixing is allowed.
Small neutrino masses are then generated by small \LN-violating parameters.
The \ISS and \LSS are prominent examples: in the \ISS, the observed masses are controlled by small \LN-violating Majorana masses \cite{Wyler:1982dd,Nandi:1985uh,Mohapatra:1986aw,Mohapatra:1986bd}, whereas in the \LSS they are controlled by suppressed \LN-violating Yukawa couplings \cite{Akhmedov:1995ip,Akhmedov:1995vm,Malinsky:2005bi}.
A characteristic consequence is that heavy Majorana neutrinos pair up into quasi-degenerate pseudo-Dirac states whose mass splittings are set by the same small symmetry-breaking parameters.
The \SPSS captures the collider-relevant limit of this class of low-scale seesaw models \cite{Antusch:2015mia,Antusch:2022ceb}.
It focuses on the regime in which a single pseudo-Dirac \HNL pair is accessible, while any additional heavy states are decoupled from collider observables.

In this work, we analyse the neutrino mixing matrix of the minimal \SPSS perturbatively in the \LN-violating parameters.
The starting point is the exact \LN-conserving Dirac limit, which is diagonalised without expanding in the active-sterile mixing.
The subsequent Takagi diagonalisation \cite{Haber:2020wco} is performed only in the small symmetry-breaking entries.
This gives analytic expressions for the light- and heavy-neutrino masses, all blocks of the neutrino mixing matrix, and the unitary defect of the active-light block.
The construction is therefore adapted to the pseudo-Dirac regime of symmetry-protected seesaws, rather than to the conventional seesaw expansion.

The central result is the symmetry-protected flavour reconstruction of the active-heavy interaction matrix.
It implements, in the single-pair \SPSS, the minimal-flavour insight of \cite{Gavela:2009cd}: the rank-two light-neutrino mass matrix fixes the normalised flavour direction of the leading \LN-conserving dimension-six operator.
At the same time, it has a purpose analogous to the \CI construction \cite{Casas:2001sr}, because it expresses active-heavy interaction matrices in terms of light-neutrino data and residual high-energy information.
The constraint being solved is, however, different.
Instead of solving the quadratic seesaw relation with a residual complex orthogonal matrix, the symmetry-protected flavour reconstruction solves the linear \LN-violating mass relation that remains after the exact Dirac limit has been diagonalised.
The residual high-energy information is reduced to a single complex light-heavy amplitude.
Its modulus fixes the overall size of the active-sterile mixing, while its phase defines a \CP-odd light-heavy invariant.
Compared with the general \CI framework and its extensions \cite{Xing:2009vb,Herrero-Garcia:2025aox,Cordero-Carrion:2018xre,Cordero-Carrion:2019qtu}, this formulation makes the physical origin and the validity range of the residual parameter manifest.

The leading phenomenological consequence is a prediction for normalised active-heavy interaction flavours.
For the pair-summed \CC interaction, the heavy-sector rotation and the overall light-heavy amplitude cancel from the flavour weight.
After neglecting active-light non-unitarity and terms suppressed by the smallness of the light-neutrino masses, the leading \NC and Higgs interactions reduce to the same flavour vector.
This vector is fixed by the measured light-neutrino masses and \PMNS parameters, up to the single Majorana phase.
It traces an ellipse in the flavour simplex, with uncertainties inherited from the global neutrino-oscillation fit \cite{Esteban:2024eli}.
This provides the symmetry-protected counterpart of the flavour-triangle analyses based on conventional \CI parametrisations \cite{Caputo:2017pit,Drewes:2018gkc,Tastet:2021vwp,Drewes:2022akb,Abada:2022wvh,Abdullahi:2022jlv}.
In the conventional description, the continuous dependence on the complex \CI angle disappears only in the quasi-degenerate, large-mixing, pair-summed limit; in the present construction the cancellation follows directly from the pseudo-Dirac structure.

The same analytic mixing matrix also makes the \NLO corrections to this \LO picture transparent.
For incoherent pair-summed flavour weights, the linear \LN-violating correction cancels.
For coherent heavy-neutrino propagation, however, the pseudo-Dirac propagation matrix turns the same \LO-\NLO structure into a time-dependent flavour interference, complementing the heavy-neutrino oscillation description of \cite{Antusch:2020pnn}.
The \NDB decay amplitude is treated in the same notation.
The light-neutrino contribution is expressed through the same Majorana-phase ellipse, while the leading heavy-neutrino correction separates into a mass-splitting term and an interference term involving the \NLO interaction block.

The organisation is as follows.
The \ISS, \LSS, and \SPSS frameworks and the notation used throughout this work are introduced in \cref{sec:protected-seesaw-models}.
In particular, the exact \LN-conserving Dirac limit and its neutrino mixing matrix are discussed in \cref{sec:Dirac limit}.
The perturbative diagonalisation of the full mass matrix and the analytic expressions for the masses and mixing matrices are derived in \cref{sec:linear Takagi}.
The symmetry-protected flavour reconstruction of the active-heavy interaction matrix, the physical light-heavy amplitude, the unitary defect of the active-light block, and the parameter count are derived in \cref{sec:light-heavy interaction matrix}.
The heavy-neutrino currents, flavour weights, NuFIT-induced likelihood regions, heavy-neutrino oscillations together with their probability conventions, and \NDB observables are discussed in \cref{sec:flavour composition}.
The main results are summarised in \cref{sec:conclusion}.
The comparison with the conventional \CI construction, the effective-operator interpretation, and the heavy-neutrino contribution to \NDB decay are given in \cref{sec:CI comparison,sec:minimal-flavour-reconstruction,sec:NLO NDB decay}.

\section{Symmetry-protected seesaw models} \label{sec:protected-seesaw-models}

Collider-detectable \TISs require light-neutrino masses to be suppressed without making the \HNLs inaccessibly heavy or the active-sterile mixing unobservably small \cite{Atre:2009rg}.
In a generic model, this suppression relies on cancellations between different contributions to the neutrino mass matrix.
Symmetry-protected models protect these cancellations through an approximately unbroken \LN-like symmetry based on a group $G_L$, where the simplest realisation is $G_L=\U(1)_L$.
The approximate symmetry keeps the light-neutrino masses small and pairs the heavy Majorana neutrinos into pseudo-Dirac states.

\subsection{\sentence\LNlong-conserving Dirac limit} \label{sec:Dirac limit}

\begin{table}
 \begin{tabular}{*4c} \toprule & $L$ & $N_1$ & $N_2$ \\\midrule
 $U(1)_L$ & $1$ & $-1$ & $1$ \\\bottomrule
 \end{tabular}
 \caption[Charges of the lepton fields under the \LNlong-like symmetry]{
 Charges of the lepton fields under the \LN-like symmetry group $G_L= U(1)_L$.
 } \label{tab:LNLS}
\end{table}

The simplest realisations of symmetry-protected seesaw models extend the \SM particle content with two sterile two-component fermions carrying opposite \LN charges.
With the charge convention in \cref{tab:LNLS}, the \LN-conserving terms are
\begin{equation} \label{eq:LN-conserving Lagrangian}
 \mathcal L_L \equiv
 i N_i^\dagger\widebar\sigma^\mu\partial_\mu N_i
 -
 \left(
 y_{1\alpha} N_1 L_\alpha \widetilde H^\dagger
 + m_D^{} N_1 N_2
 + \hc
 \right).
\end{equation}
Without loss of generality, the mass parameter $m_D^{}$ can be chosen to be real and positive.
\footnote{
While $m_D^{}$ plays the role of the Majorana mass in the standard seesaw, it is a Dirac mass in symmetry-protected seesaws.
}
After \EWSB, the Yukawa interaction and \SM Higgs \VEV generate the Dirac mass vector
\begin{align} \label{eq:Dirac mass}
\vec m_D^{}&\equiv v\vec y_1 , &
v &\approx \qty{174}{GeV} .
\end{align}
The neutrino mass matrix in the interaction basis $\row{\vec \nu, N_1, N_2}^\trans$ is
\footnote{
We indicate suppressed matrix indices using bold font.
}
\begin{equation} \label{eq:LN-conserving mass matrix}
 \mat M_L \equiv
 \begin{pmatrix}
 \mat 0 & \vec m_D^{} & \vec 0 \\
 \vec m_D^\trans & 0 & m_D^{} \\
 \vec 0^\trans & m_D^{} & 0
 \end{pmatrix}.
\end{equation}
This mass matrix has two degenerate singular values.
The five mass eigenstates $n_i$ therefore fall into two classes: three light neutrinos remain massless, and the two heavy \DOFs combine into a Dirac fermion with mass
\begin{equation} \label{eq:LN-conserving mass}
m_N^{} \equiv s m_D^{} .
\end{equation}
The singular value $s$ of the normalised mass matrix $\flatfrac{\mat M_L}{m_D^{}}$ is
\begin{equation} \label{eq:singular-value}
 s^2 \equiv 1 + \abs{\vec \theta}^2.
\end{equation}
Here the active-sterile mixing is given by the ratio of the two mass terms,
\begin{equation} \label{eq:active-sterile mixing}
 \vec \theta \equiv \flatfrac{\vec m_D^{}}{m_D^{}}.
\end{equation}
Therefore, the Lagrangian \eqref{eq:LN-conserving Lagrangian} describes a model with a single Dirac \HNL.
The physical parameters of this model are the mass \eqref{eq:LN-conserving mass} and the three flavour-mixing magnitudes contained in the active-sterile mixing \eqref{eq:active-sterile mixing}.

The Takagi matrix that factorises this mass matrix is defined by
\begin{align} \label{eq:LN conserving neutrino mixing matrix}
 \mat D_L &\equiv \mat U_L^\trans \mat M_L \mat U_L = \diag\row{0,0,0,m_N^{},m_N^{}}, &
 \mat U_L &\equiv
 \begin{pmatrix}
 \mat U_L^\nu & \mat U_L^I \\
 \mat U_L^U & \mat U_L^N
 \end{pmatrix} , &
 \mat U_L^{-1} &= \mat U_L^\dagger .
\end{align}
The matrix $\mat U_L$ is the \LN-conserving neutrino mixing matrix and is exact to all orders in the active-sterile mixing \eqref{eq:active-sterile mixing}.
The four submatrices are
\footnote{
Flavour vectors are written as column vectors, so $\inner{\vec a^\trans \vec c}$ is a bilinear scalar product, while $\tensor{\vec a\vec c^\trans}$ is a flavour-space matrix.
}
\begin{align} \label{eq:LN conserving neutrino mixing sub-matrices}
 \mat U_L^\nu &\equiv \mat 1 - \frac{\tensor{\vec \theta^\ast \vec \theta^\trans}}{s+s^2} , &
 \mat U_L^N &\equiv -i \tensor{\vec e_1 \vec e_+^\trans} + i \frac{\tensor{\vec e_2 \vec e_-^\trans}}{s} , &
 \mat U_L^I &\equiv i \frac{\tensor{\vec \theta^\ast \vec e_-^\trans}}{s} , &
 \mat U_L^U &\equiv - \frac{\tensor{\vec e_2 \vec \theta^\trans}}{s} ,
\end{align}
where the unit vectors are
\begin{align} \label{eq:unit vectors}
 \vec e_+ &\equiv \frac{\vec e_1 + i \vec e_2}{\sqrt 2} , &
 \vec e_- &\equiv \frac{\vec e_1 - i \vec e_2}{\sqrt 2} , &
 \vec e_1 &\equiv \begin{pmatrix} 1 \\ 0 \end{pmatrix} , &
 \vec e_2 &\equiv \begin{pmatrix} 0 \\ 1 \end{pmatrix} .
\end{align}
Here we have \emph{not} employed the common seesaw expansion
\begin{align} \label{eq:seesaw expansion}
\abs{\vec m_D^{}} &\ll m_D^{}, &
\vec \theta &\ll 1.
\end{align}
The active-light mixing matrix \eqref{eq:LN conserving neutrino mixing sub-matrices} is non-unitary \cite{Antusch:2006vwa,Antusch:2014woa,Aloni:2022ebm}.
Its unitary defect is
\begin{equation} \label{eq:LN conserving unitary defect}
 \mat \eta_L^\nu \equiv (\mat U_L^\nu)^\dagger \mat U_L^\nu - \mat 1 = - \frac{\tensor{\vec \theta^\ast \vec \theta^\trans}}{s^2}.
\end{equation}
The norm
\begin{equation} \label{eq:eta0norm}
 \norm{\mat \eta_L^\nu} = \sqrt{\tr (\mat \eta_L^\nu)^\dagger \mat \eta_L^\nu} = \frac{\abs{\vec \theta}^2}{s^2},
\end{equation}
provides a scalar measure of how far the active-light mixing matrix deviates from a unitary matrix.
It vanishes only if the active-sterile mixing \eqref{eq:active-sterile mixing} is absent.

\subsection{\sentence\ISSlong}

\resetacronym{ISS}

In addition to the \LN-conserving Lagrangian \eqref{eq:LN-conserving Lagrangian}, the \ISS allows one \LN-violating Majorana mass term \cite{Wyler:1982dd,Nandi:1985uh,Mohapatra:1986aw,Mohapatra:1986bd}
\begin{equation} \label{eq:LISS}
 \mathcal L_{\slashed L}^{\ISS} \equiv
 - \frac{\mu_M^{}}{2} N_2 N_2
 + \hc .
\end{equation}
At the level of the mass matrix, this gives
\begin{align} \label{eq:ISS mass matrix}
 \mat M^{\ISS} &\equiv \mat M_L + \mat M_{\slashed L}^{\ISS} , &
 \mat M_{\slashed L}^{\ISS} &\equiv
 \begin{pmatrix}
 \mat 0 & \vec 0 & \vec 0 \\
 \vec 0^\trans & 0 & 0 \\
 \vec 0^\trans & 0 & \mu_M^{}
 \end{pmatrix},
\end{align}
where $\mat M_L$ is the \LN-conserving mass matrix \eqref{eq:LN-conserving mass matrix}.
In the seesaw expansion \eqref{eq:seesaw expansion}, the light-neutrino mass matrix is
\begin{equation} \label{eq:ISS active mass}
 \widehat{\mat M}^\nu_{\ISS}
 = \mu_M^{} \tensor{\vec \theta \vec \theta^\trans} + \order{\mu_M^{}\abs{\vec \theta}^4} .
\end{equation}
Thus the small parameter $\mu_M^{}$ controls the mass of the light neutrino.
With only one sterile pair, the matrix \eqref{eq:ISS active mass} has rank one, so the minimal \ISS gives only one non-zero light mass.

\subsection{\sentence\LSSlong}

\resetacronym{LSS}

The \LSS breaks \LN through a second, suppressed Yukawa vector \cite{Akhmedov:1995ip,Akhmedov:1995vm,Malinsky:2005bi}
\begin{equation} \label{eq:LLSS}
 \mathcal L_{\slashed L}^{\text{LSS}} \equiv
 - y_{2\alpha}\tilde H^\dagger L_\alpha N_2
 + \hc .
\end{equation}
After \EWSB, the Yukawa vector leads to
\begin{equation} \label{eq:LSS mass}
 \vec \mu_D^{} \equiv v \vec y_2.
\end{equation}
The neutrino mass matrix is therefore
\begin{align} \label{eq:LSS mass matrix}
 \mat M^{\LSS} &\equiv \mat M_L + \mat M_{\slashed L}^{\LSS} , &
 \mat M_{\slashed L}^{\LSS} &\equiv
 \begin{pmatrix}
 \mat 0 & \vec 0 & \vec \mu_D^{} \\
 \vec 0^\trans & 0 & 0 \\
 \vec \mu_D^\trans & 0 & 0
 \end{pmatrix},
\end{align}
where $\mat M_L$ is the \LN-conserving mass matrix \eqref{eq:LN-conserving mass matrix}.
At \LO in the seesaw expansion \eqref{eq:seesaw expansion}, the corresponding light-neutrino mass matrix is
\begin{equation} \label{eq:LSS active mass}
 \widehat{\mat M}_{\LSS}^\nu
 =
 -\tensor{\vec \mu_D^{}\vec \theta^\trans}
 -\tensor{\vec \theta \vec \mu_D^\trans}
 + \order{\abs{\vec \mu_D^{}} \abs{\vec \theta}^2} .
\end{equation}
For one sterile pair this matrix has rank at most two.
It can therefore reproduce the two non-zero light-neutrino masses, while the overall size is protected by the smallness of $\vec \mu_D^{}$.
The phenomenology of the simplest linear-seesaw realisation, including its quasi-Dirac leptons and collider signatures, has been studied \eg in \cite{Batra:2023mds}.

\subsection{\sentence\SPSSlong} \label{sec:SPSS}

\resetacronym{SPSS}

The \ISS and \LSS are two limiting cases of the most general \LN-breaking Lagrangian.
In the \SPSS, the three possible sources of \LN violation, the \ISS term \eqref{eq:LISS}, the \LSS term \eqref{eq:LLSS}, and one additional term, are treated on equal footing \cite{Antusch:2015mia,Antusch:2022ceb}:
\begin{equation} \label{eq:LSPSS}
 \mathcal L_{\slashed L}^{\SPSS} \equiv
 \mathcal L_{\slashed L}^{\ISS} +
 \mathcal L_{\slashed L}^{\LSS} -
 \left(\frac{\mu_M^\prime}{2} N_1 N_1 + \hc \right) + \dots ,
\end{equation}
where the displayed ellipsis indicates possible interactions with additional sterile neutrinos.
For collider studies they can be assumed to decouple from the one-pair effective description.
The small breaking of $G_L$, together with the charges given in \cref{tab:LNLS}, ensures the mass hierarchy
\begin{align} \label{eq:small LN violation}
 0 & \leq \epsilon \ll 1, &
 \epsilon &\equiv \frac{\mu}{m}, &
 \mu & \in \{\abs{\mu_M^{}}, \abs{\mu_M^\prime}, \abs{\vec \mu_D^{}}\} , &
 m & \in \{\abs{\vec m_D^{}}, m_D^{}\} .
\end{align}
The symmetry does not require the small parameters to be of the same order.
The \LN-violating mass matrix that accompanies the \LN-conserving mass matrix \eqref{eq:LN-conserving mass matrix} is
\begin{align} \label{eq:LN-violating mass matrix}
 \mat M &\equiv \mat M_L + \mat M_{\slashed L}, &
 \mat M_{\slashed L} &\equiv
 \begin{pmatrix}
 \mat 0 & \vec 0 & \vec \mu_D^{} \\
 \vec 0^\trans & \mu_M^\prime & 0 \\
 \vec \mu_D^\trans & 0 & \mu_M^{}
 \end{pmatrix}.
\end{align}

\section{Perturbative \LNlong violation} \label{sec:linear Takagi}

Rotating the \LN-violating mass matrix \eqref{eq:LN-violating mass matrix} with the \LN-conserving neutrino mixing matrix \eqref{eq:LN conserving neutrino mixing matrix} into the \LN-conserving mass basis gives the mass matrix
\begin{align} \label{eq:linear rotated mass matrix}
 \widehat{\mat M}_{\slashed L} &\equiv
 \mat U_L^\trans \mat M_{\slashed L} \mat U_L =
 \begin{pmatrix}
 \widehat{\mat M}_{\slashed L}^\nu & \widehat{\mat M}_{\slashed L}^{\nu N} \\
 (\widehat{\mat M}_{\slashed L}^{\nu N})^\trans & \widehat{\mat M}_{\slashed L}^N
 \end{pmatrix} , &
 \widehat{\mat M}_{\slashed L}^\trans &= \widehat{\mat M}_{\slashed L} ,
\end{align}
where the three submatrices are
\begin{subequations} \label{eq:LNV mass matrix}
\begin{align} \label{eq:light-neutrino mass matrix}
 \widehat{\mat M}_{\slashed L}^\nu &\equiv \tensor{\vec \theta \vec \mu_\nu^\trans} + \tensor{\vec \mu_\nu \vec \theta^\trans}, &
 \vec \mu_\nu &\equiv \frac{\mu_\nu \vec \theta - \vec \mu_D^{}}{s}, &
 \mu_\nu &\equiv \frac{\mu_M^{}}{2 s} + \frac{\inner{\vec \theta^\dagger \vec \mu_D^{}}}{s+s^2} , \\ \label{eq:light-heavy mass matrix}
 \widehat{\mat M}_{\slashed L}^{\nu N} &\equiv i \tensor{\vec \mu_{\nu N}^{} \vec e_-^\trans} , &
 \vec \mu_{\nu N}^{} &\equiv - \vec \mu_\nu - \frac{\mu_N}{2} \vec \theta
, \\
 \widehat{\mat M}_{\slashed L}^N &\equiv \frac12
 \begin{pmatrix}
 -\mu_N^+ & i \mu_N^- \\
 i \mu_N^- & \mu_N^+
 \end{pmatrix} , &
 \mu_N^\pm &\equiv \mu_N \pm \mu_M^\prime , &
 \mu_N &\equiv \frac{2\inner{\vec \theta^\dagger \vec \mu_D^{}} + \mu_M^{}}{s^2} .
 \label{eq:heavy LNV mass matrix}
\end{align}
\end{subequations}
In the limits of a single \LN-violating contribution the active mass matrix becomes
\begin{align}
\eval*{\widehat{\mat M}_{\slashed L}^\nu}_{\vec \mu_D^{}=0} &= \frac{\mu_M^{}}{s^2} \tensor{\vec \theta \vec \theta^\trans} , &
\eval*{\widehat{\mat M}_{\slashed L}^\nu}_{\mu_M^{}=0} &= \tensor{\vec \theta \vec \mu_\nu^{\prime\trans}} + \tensor{\vec \mu_\nu^\prime \vec \theta^\trans} , &
\vec \mu_\nu^\prime &\equiv \frac{\mu_\nu^\prime \vec \theta - \vec \mu_D^{}}{s} , &
\mu_\nu^\prime &\equiv \frac{\inner{\vec \theta^\dagger \vec \mu_D^{}}}{s + s^2}.
\end{align}
Taking the seesaw limit \eqref{eq:seesaw expansion}, these expressions reproduce the \LO active mass matrices of the \ISS \eqref{eq:ISS active mass} and the \LSS \eqref{eq:LSS active mass}.

In the \LN-conserving mass basis, the full mass matrix is the sum of the unperturbed diagonal matrix \eqref{eq:LN conserving neutrino mixing matrix} and the perturbation \eqref{eq:linear rotated mass matrix}.
At \LO, this matrix is diagonalised by
\begin{align} \label{eq:LN-violating mixing matrix}
 \mat D &\equiv \widehat{\mat U}_{\slashed L}^\trans \widehat{\mat M} \widehat{\mat U}_{\slashed L}
 = \diag\row{m_1,m_2,m_3,m_4,m_5}, &
 \widehat{\mat U}_{\slashed L}^{-1} &= \widehat{\mat U}_{\slashed L}^\dagger , &
 \widehat{\mat M} &\equiv \mat D_L + \widehat{\mat M}_{\slashed L} .
\end{align}
We first diagonalise the active and sterile blocks separately with the diagonal blocks of the unitary matrix $\mat U_{\slashed L}$.
The remaining off-diagonal terms are then removed by a final unitary rotation $\exp \mat \Omega_{\slashed L}$, expanded to \LO in the \LN-violating parameters,
\begin{align} \label{eq:perturbative Takagi}
 \widehat{\mat U}_{\slashed L} &\equiv \mat U_{\slashed L} \exp \mat \Omega_{\slashed L} , &
 \mat U_{\slashed L}^{-1} &= \mat U_{\slashed L}^\dagger , &
 \exp \mat \Omega_{\slashed L} &= \mat 1 + \mat \Omega_{\slashed L} + \order{\epsilon^2} , &
 \mat \Omega_{\slashed L}^\dagger &= - \mat \Omega_{\slashed L} .
\end{align}

\subsection{Active-light mixing} \label{sec:active-light mixing matrix}

The active mass matrix \eqref{eq:light-neutrino mass matrix} is linear in the \LN-violating parameters.
Its Takagi matrix
\begin{align} \label{eq:light-neutrino Takagi matrix}
 \mat D^\nu_{\slashed L} &\equiv (\mat U_{\slashed L}^\nu)^\trans \widehat{\mat M}_{\slashed L}^\nu \mat U_{\slashed L}^\nu = \diag\row{m_1,m_2,m_3} , &
 (\mat U_{\slashed L}^\nu)^{-1} &= (\mat U_{\slashed L}^\nu)^\dagger ,
\end{align}
generates the \LO light-neutrino masses.
Since $\widehat{\mat M}_{\slashed L}^\nu$ has rank at most two, one of the light neutrinos remains exactly massless, while the other two acquire masses proportional to the \LN-violating mass parameters.
For \NO and \IO, the masses can be written in terms of the positive squared-mass splittings $\Delta m_{ij}^2\equiv\abs{m_i^2-m_j^2}$ as \cite{Esteban:2024eli}
\begin{subequations} \label{eq:light-neutrino masses}
\begin{align}
 \NO &: &
 m_1^2 &= 0 , &
 m_2^2 &= \Delta m_{21}^2 , &
 m_3^2 &= \Delta m_{3k}^2 , &
 k &= 1,
 \\
 \IO &: &
 m_3^2 &= 0 , &
 m_1^2 &= \Delta m_{3k}^2 - \Delta m_{21}^2, &
 m_2^2 &= \Delta m_{3k}^2 , &
 k &= 2.
\end{align}
\end{subequations}
Here $\Delta m_{21}^2$ is the smaller solar squared-mass splitting, whereas $\Delta m_{3k}^2$ denotes the larger atmospheric squared-mass splitting.
We label the massless, lighter massive, and heavier massive light-neutrino directions by
\begin{equation} \label{eq:ordering dependent light indices}
 (\one, \two, \three)\equiv
 \begin{cases}
 (1,2,3) & \text{for } \NO, \\
 (3,1,2) & \text{for } \IO .
 \end{cases}
\end{equation}
The average of the two non-zero light-neutrino masses and their mass difference are then
\begin{align} \label{eq:average light-neutrino masses}
 \ev{m_{\two\three}} &\equiv \frac{m_\two+m_\three}{2}= \abs{\vec \theta} \abs{\vec \mu_\nu} , &
 \Delta m_{\two\three} &\equiv m_\three-m_\two = 2 \abs{\inner{\vec \theta^\dagger \vec \mu_\nu}} ,
\end{align}
so that the squared mass splittings \eqref{eq:light-neutrino masses} are
\begin{subequations}
\begin{align}
 \NO &: &
 \sqrt{\Delta m_{21}^2} &= \ev{m_{\two\three}} - \frac12 \Delta m_{\two\three} , &
 \sqrt{\Delta m_{3k}^2} &= \ev{m_{\two\three}} + \frac12 \Delta m_{\two\three} ,
 \\
 \IO &: &
 \Delta m_{21}^2 &= 2 \ev{m_{\two\three}} \Delta m_{\two\three} , &
 \sqrt{\Delta m_{3k}^2} &= \ev{m_{\two\three}} + \frac12 \Delta m_{\two\three} .
\end{align}
\end{subequations}

The neutrino oscillation measurements are parameterised in terms of the unitary \PMNS matrix
\begin{align} \label{eq:PMNS matrix}
\mat U_{\PMNS} &\equiv \mat R_{23} \mat U_{13} \mat R_{12} , &
\mat U_{13} &\equiv \mat D_D^\dagger \mat R_{13} \mat D_D, &
\mat D_D &\equiv \diag\row{e^{i\delta},1,1} ,
\end{align}
where the rotation matrices are
\begin{align} \label{eq:PMNS rotations}
\mat R_{12} &\equiv
\begin{psmallmatrix}
c_{12} & s_{12} & 0\\
-s_{12} & c_{12} & 0\\
0 & 0 & 1
\end{psmallmatrix}, &
\mat R_{13} &\equiv
\begin{psmallmatrix}
c_{13} & 0 & s_{13}\\
0 & 1 & 0\\
-s_{13} & 0 & c_{13}
\end{psmallmatrix}, &
\mat R_{23} &\equiv
\begin{psmallmatrix}
1 & 0 & 0\\
0 & c_{23} & s_{23}\\
0 & -s_{23} & c_{23}
\end{psmallmatrix},
\end{align}
with $c_{ij}\equiv\cos\theta_{ij}$ and $s_{ij}\equiv\sin\theta_{ij}$.
The Dirac phase is probed by the \CP-odd invariant \cite{Jarlskog:1985ht}
\begin{equation} \label{eq:PMNS invariant}
 I_\nu \equiv \Im
 (\mat U_{\PMNS})_{e1}
 (\mat U_{\PMNS})_{\mu2}
 (\mat U_{\PMNS})_{e2}^\ast
 (\mat U_{\PMNS})_{\mu1}^\ast = c_{12}s_{12}c_{23}s_{23}c_{13}^2s_{13}
 \sin\delta.
\end{equation}
Additionally, the active-light Takagi matrix \eqref{eq:light-neutrino Takagi matrix} contains one currently unobserved Majorana phase,
\begin{align} \label{eq:Majorana phase}
\mat U_{\slashed L}^\nu &\equiv \mat U_{\PMNS} \mat D_M , &
\mat D_M &\equiv
\begin{cases}
\diag\row{1,1,e^{i\alpha}} & \text{for } \NO, \\
\diag\row{1,e^{i\alpha},1} & \text{for } \IO .
\end{cases}
\end{align}
Thus the light-neutrino sector is parameterised by the measured quantities
\begin{equation} \label{eq:light-neutrino oscillation observables}
 \vec x \equiv
 \row{
 \Delta m_{21}^2,
 \Delta m_{3k}^2,
 \theta_{12},
 \theta_{13},
 \theta_{23},
 \delta
 }^\trans ,
\end{equation}
accompanied by the Majorana phase \eqref{eq:Majorana phase}.

\subsection{Sterile-heavy mixing}

The sterile mass matrix in the \LN-conserving mass basis \eqref{eq:heavy LNV mass matrix} can be diagonalised by a real orthogonal rotation,
\begin{align} \label{eq:heavy-neutrino mixing matrix}
 \widehat{\mat D}^N &\equiv (\mat U_{\slashed L}^N)^\trans \widehat{\mat M}_{\slashed L}^N \mat U_{\slashed L}^N , &
 \mat U_{\slashed L}^N &\equiv \begin{pmatrix}
 \cos \nicefrac{\phi_N^{}}{2} & \sin \nicefrac{\phi_N^{}}{2} \\
 - \sin \nicefrac{\phi_N^{}}{2} & \cos \nicefrac{\phi_N^{}}{2}
 \end{pmatrix} , &
 (\mat U_{\slashed L}^N)^{-1} &= (\mat U_{\slashed L}^N)^\trans,
\end{align}
where the angle is given by the argument
\begin{align} \label{eq:heavy splitting phase}
\phi_N^{} &\equiv \arg \mu_+ , &
\Delta m &\equiv \abs{\mu_+} , &
\mu_\pm &\equiv \Re\mu_N^\pm - i \Im\mu_N^\mp = \mu_N^\ast \pm \mu_M^\prime ,
\end{align}
and $\mu_N^\pm$ and $\mu_N$ are defined in \eqref{eq:heavy LNV mass matrix}.
This rotation diagonalises the real part of the sterile perturbation exactly,
\begin{equation}
\Re \widehat{\mat D}^N = \frac{\Delta m}{2}
 \begin{pmatrix}
 -1 & 0 \\
 0 & 1
 \end{pmatrix} .
\end{equation}
The masses of the two heavy states $n_{\nicefrac45}$ are therefore, to \LO in the \LN-violating mass parameters,
\begin{align} \label{eq:heavy mass splitting}
 \mat D^N &\equiv \diag(m_4, m_5) , &
m_{\nicefrac45} &= m_N^{} \mp \frac12 \Delta m .
\end{align}
The remaining imaginary part is
\begin{align}
 \Im\widehat{\mat D}^N &= \frac12
 \begin{pmatrix}
 \Im\mu_-^\prime & \Re\mu_-^\prime \\
 \Re\mu_-^\prime & -\Im\mu_-^\prime
 \end{pmatrix} , &
 \mu_-^\prime \equiv \mu_- e^{-i \phi_N^{}}.
\end{align}
It is removed by a small anti-Hermitian transformation \eqref{eq:perturbative Takagi},
\begin{align} \label{eq:heavy-neutrino correction}
 \mat \Omega_{\slashed L}^N &= - \frac{i}{2} \frac{\Im \widehat{\mat D}^N}{m_N^{}} , &
 (\mat \Omega_{\slashed L}^N)^\dagger &= -\mat \Omega_{\slashed L}^N .
\end{align}

\subsection{Light-heavy mixing}

We apply the active-light Takagi matrix \eqref{eq:light-neutrino Takagi matrix} and the sterile-heavy mixing matrix \eqref{eq:heavy-neutrino mixing matrix} in the form of the block-diagonal matrix
\begin{align}
 \mat U_{\slashed L} &\equiv \begin{pmatrix}
 \mat U_{\slashed L}^\nu & \mat 0 \\
 \mat 0 & \mat U_{\slashed L}^N
 \end{pmatrix} , &
 (\mat U_{\slashed L})^{-1} &= (\mat U_{\slashed L})^\dagger ,
\end{align}
to the mass matrix in the \LN-conserving mass basis \eqref{eq:linear rotated mass matrix}.
The remaining light-heavy defect is removed by the correction
\begin{equation} \label{eq:light-heavy correction}
 m_N^{} (\mat \Omega_{\slashed L}^{\nu N})^\ast = (\mat U_{\slashed L}^\nu)^\trans \widehat{\mat M}_{\slashed L}^{\nu N} \mat U_{\slashed L}^N.
\end{equation}
Together with the heavy-sector correction \eqref{eq:heavy-neutrino correction}, this gives the full anti-Hermitian correction matrix \eqref{eq:perturbative Takagi},
\begin{equation} \label{eq:linear Takagi generator}
 \mat \Omega_{\slashed L} \equiv \begin{pmatrix}
 \mat 0 & \mat \Omega_{\slashed L}^{\nu N} \\
 -(\mat \Omega_{\slashed L}^{\nu N})^\dagger & \mat \Omega_{\slashed L}^N
 \end{pmatrix} .
\end{equation}
The complete neutrino mixing matrix for the original mass matrix \eqref{eq:LN-violating mass matrix} is
\begin{align} \label{eq:neutrino mixing matrix}
 \mat D &\equiv \mat U^\trans \mat M \mat U , &
 \mat U &\equiv \mat U_L \widehat{\mat U}_{\slashed L} , &
 \mat U^{-1} &= \mat U^\dagger , &
 \mat U &\equiv \begin{pmatrix}
 \mat U^\nu & \mat U^I \\
 \mat U^U & \mat U^N
 \end{pmatrix} ,
\end{align}
where the \LN-violating mixing matrix is defined in \eqref{eq:LN-violating mixing matrix}.
The submatrices are given to \NLO in the \LN-violating parameters.
The active-light mixing matrix is
\begin{align} \label{eq:light-neutrino mixing matrix}
 \mat U^\nu &= \mat U_0^\nu + \mat U_1^\nu + \order{\epsilon^2} , &
 \mat U_0^\nu &\equiv \mat U_L^\nu \mat U_{\slashed L}^\nu , &
 \mat U_1^\nu &\equiv -\mat U_L^I \mat U_{\slashed L}^N (\mat \Omega_{\slashed L}^{\nu N})^\dagger ,
\end{align}
where the \LN-violating active-light Takagi matrix is defined in \eqref{eq:light-neutrino Takagi matrix}.
The sterile-heavy mixing matrix is
\begin{align}
 \label{eq:mixing matrix heavy}
 \mat U^N &= \mat U_0^N + \mat U_1^N + \order{\epsilon^2} , &
 \mat U_0^N &\equiv \mat U_L^N \mat U_{\slashed L}^N ,&
 \mat U_1^N &\equiv \mat U_L^N \mat U_{\slashed L}^N \mat \Omega_{\slashed L}^N
 + \mat U_L^U \mat U_{\slashed L}^\nu \mat \Omega_{\slashed L}^{\nu N} ,
\end{align}
where the \LN-violating sterile-heavy mixing matrix and the corresponding correction matrix are defined in \eqref{eq:heavy-neutrino mixing matrix,eq:heavy-neutrino correction}.
The active-heavy interaction matrix is
\begin{align}
 \label{eq:light-heavy interaction matrix}
 \mat U^I &= \mat U_0^I + \mat U_1^I + \order{\epsilon^2} , &
 \mat U_0^I &\equiv \mat U_L^I \mat U_{\slashed L}^N , &
 \mat U_1^I &\equiv \mat U_L^I \mat U_{\slashed L}^N \mat \Omega_{\slashed L}^N
 + \mat U_L^\nu \mat U_{\slashed L}^\nu \mat \Omega_{\slashed L}^{\nu N} ,
\end{align}
where the off-diagonal correction is defined in \eqref{eq:light-heavy correction}.
The sterile-light unitary complement to the active-heavy interaction matrix is
\begin{align}
 \label{eq:unitary complement matrix}
 \mat U^U &= \mat U_0^U + \mat U_1^U + \order{\epsilon^2} , &
 \mat U_0^U &\equiv \mat U_L^U \mat U_{\slashed L}^\nu , &
 \mat U_1^U &\equiv -\mat U_L^N \mat U_{\slashed L}^N (\mat \Omega_{\slashed L}^{\nu N})^\dagger .
\end{align}
The \LN-conserving mixing matrices are given in \eqref{eq:LN conserving neutrino mixing sub-matrices}.

\section{Active-heavy interaction matrix} \label{sec:light-heavy interaction matrix}

\resetacronym{CI}

To study the interactions of \HNLs with \SM fermions in the basis of physical parameters, the active-heavy interaction matrix \eqref{eq:light-heavy interaction matrix} and its sterile-light unitary complement \eqref{eq:unitary complement matrix} must be expressed in terms of the active-light and sterile-heavy mixing matrices in \eqref{eq:light-neutrino mixing matrix,eq:mixing matrix heavy}.
The most widely used parameterisation of this type is the \CI construction for generic \TISs in the seesaw limit \cite{Casas:2001sr}.
Here we consider a construction that is closely related to the minimal-flavour seesaw construction \cite{Gavela:2009cd}, but formulated directly for the perturbative mixing matrix of the \SPSS.
For this construction we introduce a light-heavy interaction kernel, obtained by rotating the active-heavy interaction matrix \eqref{eq:light-heavy interaction matrix} into the light-neutrino mass basis.
This allows us to write the active-heavy interaction matrix and its sterile-light unitary complement as
\begin{align} \label{eq:interaction kernel}
 \mat U^I &= \mat U^\nu \widehat{\mat U}^I , &
 \mat U^U &= - \mat U^N (\widehat{\mat U}^I)^\dagger , &
 \widehat{\mat U}^I &= \widehat{\mat U}^I_0 + \widehat{\mat U}^I_1 + \order{\epsilon^2},
\end{align}
The second relation gives the sterile-light unitary complement in \eqref{eq:unitary complement matrix} and follows from the unitarity of the neutrino mixing matrix \eqref{eq:neutrino mixing matrix}.

\subsection{Light-neutrino mass basis} \label{sec:light neutrino mass basis}

At \LO, we use the active-light Takagi matrix \eqref{eq:light-neutrino Takagi matrix} to define the active-sterile mixing \eqref{eq:active-sterile mixing} and the vector that generates the light-neutrino mass matrix \eqref{eq:light-neutrino mass matrix} in the light-neutrino mass basis,
\begin{align} \label{eq:light-neutrino mass basis}
 \hat{\vec \theta} &\equiv (\mat U_{\slashed L}^\nu)^\trans \vec \theta , &
 \hat{\vec \mu}_\nu &\equiv (\mat U_{\slashed L}^\nu)^\trans \vec \mu_\nu .
\end{align}
Components in this basis are labelled according to \eqref{eq:ordering dependent light indices}.
The massless direction implies
\begin{align}
(\hat{\vec \theta})_\one &= 0 , &
(\hat{\vec \mu}_\nu)_\one &= 0 .
\end{align}
The \LN-conserving active-heavy interaction matrix \eqref{eq:LN conserving neutrino mixing sub-matrices} can be rewritten in terms of the \LO active-light mixing matrix \eqref{eq:light-neutrino mixing matrix},
\begin{align}
 \mat U_L^I &= \frac{i}{s} \tensor{\vec \theta^\ast \vec e_-^\trans}
 = i \mat U_0^\nu
 \tensor{\hat{\vec \theta}{}^\ast \vec e_-^\trans} , &
\mat U_0^\nu \hat{\vec \theta}{}^\ast &= \frac{\vec \theta^\ast}{s}.
\end{align}
Therefore, the \LO interaction kernel \eqref{eq:interaction kernel} becomes
\begin{align} \label{eq:heavy-neutrino direction}
 \widehat{\mat U}^I_0 &= i \tensor{\hat{\vec \theta}{}^\ast (\vec d_{\slashed L}^N)^\trans} , &
 \vec d_{\slashed L}^N &\equiv (\mat U_{\slashed L}^N)^\trans \vec e_- = \vec e_- e^{\flatfrac{i\phi_N^{}}{2}} ,
\end{align}
where we have introduced the heavy direction selected by the \LN-violating sterile-heavy Takagi rotation \eqref{eq:heavy-neutrino mixing matrix}.
In the light-neutrino mass basis, the light-neutrino mass matrix \eqref{eq:light-neutrino mass matrix}
\begin{align} \label{eq:light-neutrino mass basis constraint}
 \mat D^\nu_{\slashed L} &= \tensor{\hat{\vec \theta} \hat{\vec \mu}_\nu^\trans}
 + \tensor{\hat{\vec \mu}_\nu \hat{\vec \theta}^\trans}
\end{align}
is diagonal.
This constraint fixes the symmetric product of $\hat{\vec \theta}$ and $\hat{\vec \mu}_\nu$ and is therefore invariant under an opposite rescaling of the two vectors.
This rescaling freedom can be expressed by introducing the Takagi direction
\begin{align} \label{eq:Takagi sign}
 \hat{\vec d}^\sigma &\equiv \row{0, 1, \sigma i}^\trans , &
 \sigma &\in\{\pm1\} ,
\end{align}
where the Takagi sign labels the two allowed relative phases between the two massive light-neutrino directions.
The light-neutrino direction in the light-neutrino mass basis is then
\begin{align} \label{eq:light-neutrino direction in the light mass basis}
 \hat{\vec d}^\nu
 &\equiv
 \frac{(\mat D^\nu_{\slashed L})^{\nicefrac12} \hat{\vec d}^\sigma}{\sqrt{2\ev{m_{\two\three}}}} =
 \frac{\row{0, \sqrt{m_\two}, \sigma i\sqrt{m_\three}}^\trans}{\sqrt{m_\two+m_\three}} , &
\inner{(\hat{\vec d}^\nu)^\dagger\hat{\vec d}^\nu} &= 1 ,
\end{align}
where the average light-neutrino mass is defined in \eqref{eq:average light-neutrino masses}.
The rescaling freedom of \eqref{eq:light-neutrino mass basis constraint} can then be parametrised by a complex parameter $\vartheta_0$,
\begin{align} \label{eq:rescaling parameterisation}
 \hat{\vec \theta} &= \vartheta_0 \hat{\vec d}^\nu, &
 \hat{\vec \mu}_\nu &= \frac{\ev{m_{\two\three}}}{ \vartheta_0}(\hat{\vec d}^\nu)^\ast .
\end{align}
Using the light-neutrino direction \eqref{eq:light-neutrino direction in the light mass basis}, the complex rescaling parameter and the Takagi sign \eqref{eq:Takagi sign} can be expressed in terms of the active-sterile mixing in the light-neutrino mass basis \eqref{eq:light-neutrino mass basis},
\begin{align} \label{eq:light-heavy amplitude derivation}
 \vartheta_0 &= \inner{(\hat{\vec d}^\nu)^\dagger\hat{\vec \theta}} = \frac{\ev{m_{\two\three}}}{\inner{(\hat{\vec d}^\nu)^\trans \hat{\vec \mu}_\nu}}, &
 \sigma &= -i
 \left(\frac{(\hat{\vec \theta})_\two}{\sqrt{m_\two}}\right)^{-1}
 \frac{(\hat{\vec \theta})_\three}{\sqrt{m_\three}}
 .
\end{align}
The phase of the complex rescaling parameter defines the light-sector phase,
\begin{align} \label{eq:light-sector phase}
 \arg \vartheta_0
 &=
 \frac12
 \arg\left(-\sigma i\Phi_\nu\right)
 \bmod \pi,
 &
 -\sigma i\Phi_\nu
 &=
 e^{2i\arg \vartheta_0},
\end{align}
where the phase factor is built from the only phase-sensitive off-diagonal product in the light-neutrino mass basis
\begin{align} \label{eq:light-sector phase factor}
 \Phi_\nu
 &\equiv
 \frac{\mu_I^{}}{\abs{\mu_I^{}}}, &
 \mu_I^{}
 &\equiv
 (\tensor{\hat{\vec \theta}\hat{\vec \mu}_\nu^\dagger})_{\two\three}
 =
 (\tensor{\hat{\vec \theta}\hat{\vec \mu}_\nu^\dagger})_{\three\two}
 .
\end{align}
The heavy direction in \eqref{eq:heavy-neutrino direction} contributes a factor of $\phi_N^{}/2$ to the phase.
Using the definitions in \eqref{eq:heavy splitting phase}, the corresponding heavy-sector phase factor is
\begin{align} \label{eq:heavy-sector phase factor}
 \Phi_N
 &\equiv
 \frac{\mu_+}{\abs{\mu_+}}
 =
 e^{i\phi_N^{}} .
\end{align}
The physical light-heavy amplitude is therefore
\begin{align} \label{eq:light-heavy phase convention}
 \vartheta &\equiv -i\vartheta_0 e^{-i\phi_N^{}/2}.
\end{align}
Its absolute value is given by the norm of the active-sterile mixing \eqref{eq:active-sterile mixing}, while its phase is the physical combination of the light-sector phase \eqref{eq:light-sector phase} and the heavy-sector phase \eqref{eq:heavy splitting phase},
\begin{align} \label{eq:light-heavy amplitude}
 \abs{\vartheta} &= \abs{ \vartheta_0} = \abs{\vec \theta} , &
 \arg \vartheta &= \arg \vartheta_0 - \frac{\phi_N^{}}{2} - \frac{\pi}{2} \bmod \pi .
\end{align}
The phase can be extracted from a phase factor that combines the Takagi sign \eqref{eq:Takagi sign}, the light-sector phase factor \eqref{eq:light-sector phase factor}, and the heavy-sector phase factor \eqref{eq:heavy-sector phase factor}:
\begin{align} \label{eq:light-heavy invariant}
 2\arg \vartheta &= \arg \Phi_{\nu N}\bmod 2\pi , &
 \Phi_{\nu N}
 &\equiv
 \sigma i \Phi_\nu\Phi_N^\ast
 =
 e^{2i\arg \vartheta} .
\end{align}
The phase factor can be written directly in terms of the Lagrangian parameters
\begin{align} \label{eq:light-heavy invariant Lagrangian}
 \Phi_{\nu N}
 &=
 \frac{\mu_\Phi^\ast}{\abs{\mu_\Phi^{}}} , &
 \mu_\Phi^{}
 &\equiv
 \mu_+ \inner{\vec \theta^\dagger\vec \mu_\nu}.
\end{align}
Here the active-sterile mixing, the neutrino-mass generating vector, and the mass-splitting parameter are defined in \eqref{eq:active-sterile mixing}, \eqref{eq:light-neutrino mass matrix}, and \eqref{eq:heavy splitting phase}, respectively.
A pair of Jarlskog-like \CP-even and -odd invariants is then
\begin{align} \label{eq:light-heavy Jarlskog invariants}
 R_{\nu N}
 &=
 \Re\Phi_{\nu N}
 =
 \cos 2\arg\vartheta
 =
 \frac{\vartheta^2+(\vartheta^\ast)^2}{2\abs{\vartheta}^2} ,
 &
 I_{\nu N}
 &=
 \Im\Phi_{\nu N}
 =
 \sin 2\arg\vartheta
 =
 \frac{\vartheta^2-(\vartheta^\ast)^2}{2i\abs{\vartheta}^2}.
\end{align}

\subsection{Active-heavy interaction matrices} \label{sec:heavy-neutrino interaction matrices}

Eliminating the active-sterile mixing \eqref{eq:active-sterile mixing} in the interaction kernel \eqref{eq:heavy-neutrino direction} with the parametrisation \eqref{eq:rescaling parameterisation} results in
\begin{align} \label{eq:LO interaction kernel}
 \widehat{\mat U}^I_0 &= \vartheta^\ast
 \tensor{(\hat{\vec d}^\nu)^\ast \vec e_-^\trans}.
\end{align}
Using the relation to the interaction kernel \eqref{eq:interaction kernel}, the active-heavy interaction matrix \eqref{eq:light-heavy interaction matrix} is
\begin{align} \label{eq:LO light-heavy interaction matrix}
\mat U_0^I &= \tensor{\vec c_0^I \vec e_-^\trans} , &
\vec c_0^I &= \frac{\vartheta^\ast}{s} \vec d_{\slashed L}^\theta ,
\end{align}
where $\vec d_0^\theta$ is the active-sterile mixing direction and $\vec d_0^\mu$ is the corresponding neutrino-mass-generation direction,
\begin{align} \label{eq:light-neutrino interaction directions}
 \vec d_0^\theta
 &\equiv
 \mat U_0^\nu(\hat{\vec d}^\nu)^\ast
 =
 \mat U_L^\nu\vec d_{\slashed L}^\theta
 =
 \frac1s \vec d_{\slashed L}^\theta ,
 &
 \vec d_0^\mu
 &\equiv
 \mat U_0^\nu\hat{\vec d}^\nu
 =
 \mat U_L^\nu\vec d_{\slashed L}^\mu
 =
 \left(
 \mat 1
 -
 \frac{\tensor{\vec \theta^\ast\vec \theta^\trans}}{s+s^2}
 \right)
 \vec d_{\slashed L}^\mu ,
\end{align}
where the last equalities use \eqref{eq:light-neutrino mass basis,eq:rescaling parameterisation} and $\mat U_0^\nu=\mat U_L^\nu\mat U_{\slashed L}^\nu$.
The corresponding \LN-violating components are
\begin{align} \label{eq:light-neutrino direction}
 \vec d_{\slashed L}^\theta &\equiv \mat U_{\slashed L}^\nu (\hat{\vec d}^\nu)^\ast ,
 &
 \vec d_{\slashed L}^\mu &\equiv \mat U_{\slashed L}^\nu \hat{\vec d}^\nu .
\end{align}
The sterile-light unitary complement \eqref{eq:unitary complement matrix} is
\begin{align} \label{eq:UU compact}
 \mat U_0^U &=
 -\frac{\vartheta_0}{s}
 \tensor{\vec e_2(\hat{\vec d}^\nu)^\trans}
 =
 - i \frac{\vartheta e^{i\phi_N^{}/2}}{s}
 \tensor{\vec e_2(\hat{\vec d}^\nu)^\trans}.
\end{align}
In the last line we used the phase convention \eqref{eq:light-heavy phase convention}.
The explicit phase in the unitary complement is convention dependent and cancels when the same expression is written in terms of $\vartheta_0$.
Thus, at \LO, the active-heavy interaction matrix \eqref{eq:light-heavy interaction matrix} and its sterile-light unitary complement \eqref{eq:unitary complement matrix} are fixed by the light-neutrino masses \eqref{eq:light-neutrino masses}, the active-light Takagi matrix \eqref{eq:light-neutrino Takagi matrix}, the physical light-heavy amplitude \eqref{eq:light-heavy amplitude}, and the Takagi sign \eqref{eq:Takagi sign}.

The \NLO contribution to the interaction kernel is
\begin{align} \label{eq:NLO interaction kernel}
 \widehat{\mat U}^I_1
 &=
 \mat\Omega_{\slashed L}^{\nu N}
 +
 \widehat{\mat U}^I_0
 \mat\Omega_{\slashed L}^{N}
 +
 \widehat{\mat U}^I_0
 (\mat\Omega_{\slashed L}^{\nu N})^\dagger
 \widehat{\mat U}^I_0 .
\end{align}
It is proportional to the correction matrices \eqref{eq:light-heavy correction,eq:heavy-neutrino correction} and depends on the \LO interaction kernel \eqref{eq:LO interaction kernel}.
The light-heavy correction \eqref{eq:light-heavy correction} can be expressed in terms of the heavy direction \eqref{eq:heavy-neutrino direction},
\begin{align} \label{eq:NLO light-heavy correction}
 \mat\Omega_{\slashed L}^{\nu N}
 =
 -\frac{i}{m_N^{}}
 \hat{\vec \mu}_{\nu N}^\ast
 (\vec d_{\slashed L}^{N})^\dagger =
 \frac{1}{m_N^{}}
 \left[
 \frac{\ev{m_{\two\three}}}{\vartheta^\ast} \hat{\vec d}^\nu
 - \frac{\mu_N^\ast e^{-i\phi_N}}{2} \vartheta^\ast (\hat{\vec d}^\nu)^\ast
 \right]
 \vec e_+^\trans ,
\end{align}
where we have introduced the light-heavy mass matrix generating vector \eqref{eq:light-heavy mass matrix} in the neutrino mass basis
\begin{align}
\hat{\vec \mu}_{\nu N} &\equiv (\mat U_{\slashed L}^\nu)^\trans \vec \mu_{\nu N} .
\end{align}
The last term of the interaction kernel \eqref{eq:NLO interaction kernel} is identically zero, and the other contributions simplify to
\begin{align}
 \widehat{\mat U}^I_1
 &=
 \frac{1}{m_N^{}}
 \left[
 \frac{\Delta m}{4} \vartheta^\ast (\hat{\vec d}^\nu)^\ast
 - \frac{\ev{m_{\two\three}}}{\vartheta^\ast}\hat{\vec d}^\nu
 \right]
\vec e_+^\trans ,
 \label{eq:NLO light-heavy interaction matrix compact}
\end{align}
where only physical parameters and the direction vector remain.
The corresponding \NLO contribution to the active-heavy interaction matrix \eqref{eq:light-heavy interaction matrix} follows from
\begin{align}
 \mat U^I_1
 &=
 \mat U^\nu_0
 \widehat{\mat U}^I_1
 +
 \mat U^\nu_1
 \widehat{\mat U}^I_0 , &
 \mat U_1^\nu
 \widehat{\mat U}_0^I
 &=
 -
 \mat U_0^\nu
 \widehat{\mat U}_0^I
 (\mat\Omega_{\slashed L}^{\nu N})^\dagger
 \widehat{\mat U}_0^I
 =
 \mat 0 .
\end{align}
Using \eqref{eq:NLO light-heavy interaction matrix compact} and the directions in \eqref{eq:light-neutrino interaction directions}, one obtains
\begin{align} \label{eq:NLO light-heavy interaction matrix}
 \mat U^I_1
 &= \tensor{\vec c_1^I \vec e_+^\trans} , &
 \vec c^I_1
 &=
 \frac{\vartheta^\ast}{4s}
 \frac{\Delta m}{m_N^{}}
 \vec d_{\slashed L}^\theta
 -
 \frac{1}{\vartheta^\ast}
 \frac{\ev{m_{\two\three}}}{m_N^{}}
 \vec d_0^\mu .
\end{align}
The sterile-light unitary complement can be obtained directly from \eqref{eq:unitary complement matrix}.
With
\begin{align}
\mat U_{\slashed L}^N\vec e_- &= e^{-i\phi_N^{}/2}\vec e_-, &
\mat U_L^N\vec e_-&=-i\vec e_1,
\end{align}
the \NLO contribution is
\begin{align} \label{eq:NLO unitary complement matrix compact}
 \mat U^U_1
 &=
 \frac{i}{m_N^{}}
 \tensor{
 \vec e_1
 \left[
 \frac{\ev{m_{\two\three}}}{\vartheta}
 e^{-i\phi_N^{}/2}
 (\hat{\vec d}^\nu)^\dagger
 -
 \frac{\mu_N^{}\vartheta}{2}
 e^{i\phi_N^{}/2}
 (\hat{\vec d}^\nu)^\trans
 \right]
 } ,
\end{align}
Equivalently, the same result follows from the unitary relation
\begin{align}
 \mat U^U_1
 &=
 -
 \mat U^N_0
 (\widehat{\mat U}^I_1)^\dagger
 -
 \mat U^N_1
 (\widehat{\mat U}^I_0)^\dagger ,
\end{align}
where the \LO term of the active-heavy interaction matrix, the active-light mixing matrix, and the sterile-heavy mixing matrix are given in \eqref{eq:LO light-heavy interaction matrix}, \eqref{eq:light-neutrino mixing matrix}, and \eqref{eq:heavy-neutrino mixing matrix}, respectively.

\subsection{Non-unitarity of the active-light mixing matrix}

At \LO, the active-light mixing matrix \eqref{eq:light-neutrino mixing matrix} factorises into two parts.
The \LN-violating factor is the unitary \PMNS matrix measured in neutrino oscillation experiments, augmented by one Majorana phase \eqref{eq:Majorana phase}.
The \LN-conserving factor carries the non-unitarity inherited from the Dirac limit.
The corresponding unitary defect is a unitary rotation of the defect \eqref{eq:LN conserving unitary defect},
\begin{equation} \label{eq:unitarity defect}
\mat \eta^\nu \equiv (\mat U^\nu)^\dagger \mat U^\nu - \mat 1 = (\mat U_{\slashed L}^\nu)^\dagger \mat \eta_L^\nu \mat U_{\slashed L}^\nu + \order{\epsilon^2} ,
\end{equation}
where the linear terms vanish because $(\mat U_L^U)^\dagger \mat U_L^N (\widehat{\mat M}_{\slashed L}^{\nu N})^\trans = \mat 0$.
In the light-neutrino mass basis its form
\begin{align} \label{eq:LNV unitary defect hatted}
 \mat \eta^\nu
 =
 -\frac{\tensor{\hat{\vec \theta}{}^\ast\hat{\vec \theta}^\trans}}{s^2}
 +\order{\epsilon^2} ,
\end{align}
resembles the unitary defect in the \LN-conserving limit \eqref{eq:LN conserving unitary defect}.
Using the parametrisation \eqref{eq:light-neutrino direction in the light mass basis} gives
\begin{align} \label{eq:LNV unitary defect physical}
 \mat \eta^\nu &=
 -\frac{\abs{\vartheta}^2}{1+\abs{\vartheta}^2}
 \tensor{(\hat{\vec d}^\nu)^\ast(\hat{\vec d}^\nu)^\trans}
 +\order{\epsilon^2} , &
 s^2&=1+\abs{\vartheta}^2.
\end{align}
Here we used the singular value \eqref{eq:singular-value} and the light-heavy amplitude \eqref{eq:light-heavy amplitude}.
Using the explicit expression \eqref{eq:light-neutrino direction in the light mass basis}, this can be written as
\begin{align} \label{eq:LNV unitary defect explicit}
 \mat \eta^\nu
 =
 -\frac{\abs{\vartheta}^2}{1+\abs{\vartheta}^2}
 \frac{1}{m_\two+m_\three}
 \begin{pmatrix}
 0 & 0 & 0 \\
 0 & m_\two & \sigma i\sqrt{m_\two m_\three} \\
 0 & -\sigma i\sqrt{m_\two m_\three} & m_\three
 \end{pmatrix}
 +\order{\epsilon^2}.
\end{align}
The scalar measure of the unitary defect
\begin{equation} \label{eq:LNV non-unitarity measure}
 \norm{\mat \eta^\nu} = \norm{\mat \eta_L^\nu} + \order{\epsilon^2} = \frac{\abs{\vartheta}^2}{1+\abs{\vartheta}^2} + \order{\epsilon^2},
\end{equation}
is therefore unchanged relative to the \LN-conserving limit \eqref{eq:eta0norm}.

\subsection{Physical model parameters} \label{sec:model parameters}

After the charged-lepton masses have been fixed, the continuous physical parameters of the minimal \SPSS can be organised into light-sector, heavy-sector, and light-heavy-mixing parameters.

\begin{description}

\item[Light-sector parameters]

The rank-two light-neutrino mass matrix fixes two non-zero masses, parameterised by the mass-squared differences in \eqref{eq:light-neutrino masses}, and the unitary matrix \eqref{eq:Majorana phase}.
The latter contains the three measured \PMNS angles, the Dirac phase probed by \eqref{eq:PMNS invariant}, and one Majorana phase.

\item[Heavy-sector parameters]

The pseudo-Dirac heavy spectrum is parameterised by the \LN-conserving mass \eqref{eq:LN-conserving mass} and the mass splitting \eqref{eq:heavy splitting phase}.

\item[Mixing parameters]

The modulus of the complex amplitude \eqref{eq:light-heavy amplitude} gives the size of the active-sterile mixing, $\abs{\vartheta}=\abs{\vec \theta}$, while its phase can be determined by the light-heavy \CP-odd invariant \eqref{eq:light-heavy Jarlskog invariants}.
The Takagi sign \eqref{eq:Takagi sign} is a discrete branch label rather than a continuous parameter; over the full Majorana-phase range it can be absorbed as in \eqref{eq:Takagi sign phase shift}.

\end{description}

The count agrees with the two-singlet \TIS,
\begin{align} \label{eq:SPSS parameter count}
 N_{\SPSS}^{\text{cont}}
 &=
 \smallunderbrace{2}_{m_\two,m_\three}
 +
 \smallunderbrace{(3+1+1)}_{\PMNS}
 +
 \smallunderbrace{2}_{m_N^{},\Delta m}
 +
 \smallunderbrace{2}_{\vartheta}
 =
 11 .
\end{align}

\section{Flavour composition of active-heavy interactions} \label{sec:flavour composition}

The neutrino mixing matrix \eqref{eq:neutrino mixing matrix} relates neutrino interaction and mass eigenstates by
\begin{equation}
 \begin{pmatrix}
 \vec \nu \\
 \vec N
 \end{pmatrix} = \begin{pmatrix}
 \mat U^\nu & \mat U^I \\
 \mat U^U & \mat U^N
 \end{pmatrix}
 \begin{pmatrix}
 \vec n_\nu \\
 \vec n_N^{}
 \end{pmatrix},
\end{equation}
where $\vec n_\nu$ contains the three light neutrinos and $\vec n_N^{} \equiv \row{n_4, n_5}^\trans$ contains the heavy pseudo-Dirac pair.
This matrix enters the currents that describe neutrino interactions with gauge and Higgs bosons after rotating to the neutrino mass basis \cite{Schechter:1980gr, Atre:2009rg}.

\subsection{Heavy-neutrino currents} \label{sec:heavy currents}

\resetacronym{CC}
\resetacronym{NC}

The \CC interaction depends only on the upper part of the neutrino mixing matrix \eqref{eq:neutrino mixing matrix},
\begin{equation} \label{eq:CC matrix}
 \widehat{\mat W} \equiv \row{\mat U^\nu, \mat U^I}.
\end{equation}
In the light-neutrino mass basis, the \LO submatrices are
\begin{align} \label{eq:charge current LO sub-matrices}
\widehat{\mat W}^\nu_0
&=
\mat U_0^\nu
=
\left(
 \mat 1
-
 \frac{\abs{\vartheta}^2}{s+s^2}
 \tensor{\vec d_{\slashed L}^\theta(\vec d_{\slashed L}^\theta)^\dagger}
 \right)
 \mat U_{\PMNS}\mat D_M ,
&
\widehat{\mat W}^I_0
&=
\mat U_0^I
=
\frac{\vartheta^\ast}{s}
\tensor{\vec d_{\slashed L}^\theta \vec e_-^\trans}.
\end{align}
The corresponding \CC interaction is
\begin{align} \label{eq:charged current mass basis}
 \mathcal L_{\CC}^N &\equiv -\frac{g}{\sqrt2}
 W^-_\mu
 J_W^\mu
 + \hc , &
 J_W^\mu &\equiv \vec \ell^\dagger\widebar\sigma^\mu
 \left(
 \mat U^\nu \vec n_\nu
 +
 \mat U^I \vec n_N^{}
 \right).
\end{align}
The \NC interaction depends on the square of the matrix \eqref{eq:CC matrix},
\begin{align} \label{eq:NC matrix}
 \widehat{\mat Z} &\equiv \widehat{\mat W}^\dagger \widehat{\mat W} = \begin{pmatrix}
 \widehat{\mat Z}^\nu & \widehat{\mat Z}^{\nu N} \\
 (\widehat{\mat Z}^{\nu N})^\dagger & \widehat{\mat Z}^N
 \end{pmatrix}, &
 \widehat{\mat Z}^\dagger &= \widehat{\mat Z},
\end{align}
where the submatrices are
\begin{align} \label{eq:neutral current sub-matrices}
 \widehat{\mat Z}^\nu &\equiv (\mat U^\nu)^\dagger \mat U^\nu, &
 \widehat{\mat Z}^{\nu N} &\equiv (\mat U^\nu)^\dagger \mat U^I, &
 \widehat{\mat Z}^N &\equiv (\mat U^I)^\dagger \mat U^I .
\end{align}
In the light-neutrino mass basis one finds to \LO
\begin{align} \label{eq:neutral current LO sub-matrices}
\widehat{\mat Z}^\nu_0
&=
\mat 1
-
\frac{\abs{\vartheta}^2}{s^2}
\tensor{(\hat{\vec d}^\nu)^\ast(\hat{\vec d}^\nu)^\trans},
&
\widehat{\mat Z}^{\nu N}_0
&=
\frac{\vartheta^\ast}{s^2}
\tensor{(\hat{\vec d}^\nu)^\ast\vec e_-^\trans},
&
\widehat{\mat Z}^N_0
&=
\frac{\abs{\vartheta}^2}{s^2}
\tensor{\vec e_-^\ast\vec e_-^\trans}.
\end{align}
Using the definitions of the unit vectors \eqref{eq:unit vectors} and the light-neutrino direction \eqref{eq:light-neutrino direction in the light mass basis}, the same matrices in the ordered basis \eqref{eq:ordering dependent light indices} are
\begin{subequations} \label{eq:neutral current LO sub-matrices explicit}
\begin{align}
\widehat{\mat Z}^\nu_0
&=
\mat 1
-
\frac{\abs{\vartheta}^2}{s^2(m_\two+m_\three)}
\begin{pmatrix}
0 & 0 & 0\\
0 & m_\two
& \sigma i \sqrt{m_\two m_\three}
\\
0 & - \sigma i \sqrt{m_\two m_\three}
& m_\three
\end{pmatrix}
,
\\
\widehat{\mat Z}^{\nu N}_0
&=
\frac{\vartheta^\ast}
{s^2\sqrt{2(m_\two+m_\three)}}
\begin{pmatrix}
0 & 0\\
\sqrt{m_\two} & -i\sqrt{m_\two}\\
-\sigma i\sqrt{m_\three} & -\sigma\sqrt{m_\three}
\end{pmatrix}, &
\widehat{\mat Z}^N_0
&=
\frac{\abs{\vartheta}^2}{2s^2}
\begin{pmatrix}
1 & -i\\
i & 1
\end{pmatrix}.
\end{align}
\end{subequations}
The heavy-neutrino part of the $Z$-boson current is
\begin{align} \label{eq:Z current heavy}
 \mathcal L_{\NC}^N &\equiv
 -\frac{g}{2\cos\theta_W^{}}
 Z_\mu
 J_Z^\mu , &
 J_Z^\mu &\equiv \vec n_\nu^\dagger\widebar\sigma^\mu
 \widehat{\mat Z}^{\nu N}
 \vec n_N^{}
 +
 \vec n_N^\dagger\widebar\sigma^\mu
 (\widehat{\mat Z}^{\nu N})^\dagger
 \vec n_\nu
 +
 \vec n_N^\dagger\widebar\sigma^\mu
 \widehat{\mat Z}^N
 \vec n_N^{} .
\end{align}
The Higgs interactions depend on the combination of the diagonal neutrino mass matrix \eqref{eq:neutrino mixing matrix} and the \NC matrix \eqref{eq:NC matrix}
\begin{align} \label{eq:Higgs coupling matrix}
 \widehat{\mat H} &\equiv \mat D \widehat{\mat Z} + \widehat{\mat Z}^\ast \mat D , &
 \widehat{\mat H}^\trans &= \widehat{\mat H} ,
\end{align}
whose components are
\begin{equation} \label{eq:Higgs coupling sub-matrices}
\begin{aligned}
\widehat{\mat H}^\nu
&\equiv
\mat D^\nu_{\slashed L}\widehat{\mat Z}^\nu
+(\widehat{\mat Z}^\nu)^\ast\mat D^\nu_{\slashed L},
&
\widehat{\mat H}^{\nu N}
&\equiv
\mat D^\nu_{\slashed L}\widehat{\mat Z}^{\nu N}
+(\widehat{\mat Z}^{\nu N})^\ast\mat D^N,
\\
\widehat{\mat H}^N
&\equiv
\mat D^N\widehat{\mat Z}^N
+(\widehat{\mat Z}^N)^\ast\mat D^N.
\end{aligned}
\end{equation}
At \LO the explicit expressions are
\begin{subequations} \label{eq:Higgs coupling LO sub-matrices explicit}
\begin{align}
\widehat{\mat H}^\nu_0
&= 2
\begin{pmatrix}
0 & 0 & 0 \\
0 & m_\two & 0 \\
0 & 0 & m_\three
\end{pmatrix} +
\frac{\abs{\vartheta}^2}{s^2(m_\two+m_\three)}
\begin{pmatrix}
0 & 0 & 0\\
0 & -2 m_\two^2
& \sigma i \Delta m_{\two\three} \sqrt{m_\two m_\three}
\\
0 & \sigma i \Delta m_{\two\three} \sqrt{m_\two m_\three}
& -2 m_\three^2
\end{pmatrix},
\\
\widehat{\mat H}^{\nu N}_0
&=
\frac{1}{s^2\sqrt{2(m_\two+m_\three)}}
\begin{pmatrix}
0 & 0\\
\sqrt{m_\two}(\vartheta m_4+\vartheta^\ast m_\two) &
i\sqrt{m_\two}(\vartheta m_5-\vartheta^\ast m_\two) \\
i\sigma\sqrt{m_\three}(\vartheta m_4-\vartheta^\ast m_\three) &
-\sigma\sqrt{m_\three}(\vartheta m_5+\vartheta^\ast m_\three)
\end{pmatrix},
\\
\widehat{\mat H}^N_0
&=
\frac{\abs{\vartheta}^2}{2s^2}
\begin{pmatrix}
2m_4 & i\Delta m\\
i\Delta m & 2m_5
\end{pmatrix}.
\end{align}
\end{subequations}
The heavy-neutrino Higgs interactions read
\begin{align} \label{eq:Higgs current heavy}
 \mathcal L_{NH} &\equiv
 -\frac{h}{2v}
 J_h
 + \hc , &
 J_h &\equiv \vec n_\nu^\trans
 \widehat{\mat H}^{\nu N}
 \vec n_N^{}
 +
 \vec n_N^\trans
 (\widehat{\mat H}^{\nu N})^\trans
 \vec n_\nu
 +
 \vec n_N^\trans
 \widehat{\mat H}^N
 \vec n_N^{} .
\end{align}
Consequently, the \CC interaction and the leading light-heavy parts of the \NC and Higgs interactions probe the same flavour direction given by the active flavour direction \eqref{eq:light-neutrino direction} fixed by neutrino oscillation observables and the Majorana phase.
The light-heavy amplitude \eqref{eq:light-heavy amplitude} enters only as an overall normalisation.

\subsection{Active-heavy flavour weights} \label{sec:active-heavy flavour weights}

The interaction matrices derived in \cref{sec:heavy currents} carry interaction flavour in their rows.
For such a matrix or flavour vector, the flavour weight
\begin{align} \label{eq:weight map}
 \weight x &\equiv \frac{\main x x^\dagger}{\tr x x^\dagger} , &
 (\weight x)_\alpha &\geq 0 , &
 \sum_{\alpha=e,\mu,\tau}
 (\weight x)_\alpha &= 1 ,
\end{align}
defines a point in the flavour simplex.
Here $\main$ extracts the main diagonal of a matrix.

Using the active flavour direction \eqref{eq:light-neutrino direction}, the \CC interaction \eqref{eq:charged current mass basis} of a single heavy state can be written as
\begin{equation}
 \mat U_0^I \vec e_a = \frac{\vartheta^\ast}{s}
 (\vec e_-)_a
 \vec d_{\slashed L}^\theta .
\end{equation}
Here $\vec e_a$ is the two-component unit vector \eqref{eq:unit vectors} that selects the heavy mass eigenstate $n_{3+a}$.
The \CC flavour point for this individual heavy state is therefore
\begin{align}
 \vec w_{0,a}^{\CC} &\equiv \weight(\mat U_0^I\vec e_a) = \weight(\vec d_{\slashed L}^\theta),
\end{align}
where the second equality follows because scalar prefactors cancel in the $\weight$ map \eqref{eq:weight map}.
The inclusive point obtained after summing over the pseudo-Dirac pair is identical,
\begin{align} \label{eq:CC weight vector}
 \vec w^{\CC}_0 &\equiv \weight(\mat U_0^I) = \weight(\vec d_{\slashed L}^\theta),
\end{align}
because $\abs{\vec e_-}^2=1$.
Thus the \CC flavour weight depends only on the active flavour direction \eqref{eq:light-neutrino direction}.
The dependence on the light-heavy amplitude \eqref{eq:light-heavy amplitude} cancels from the flavour weights.
The components of the \CC flavour weight vector are
\begin{align} \label{eq:LO CC weight vector}
 (\vec w^{\CC}_0)_\alpha
 &=
 \frac{
 \abs{
 \sqrt{m_\two}(\mat U_{\slashed L}^\nu)_{\alpha \two}
 - \sigma i
 \sqrt{m_\three}(\mat U_{\slashed L}^\nu)_{\alpha \three}
 }^2
 }{m_\two+m_\three} ,
\end{align}
where the indices follow the ordering convention in \eqref{eq:ordering dependent light indices}.

To connect the \NC interaction matrix \eqref{eq:neutral current sub-matrices} and the Higgs interaction matrix \eqref{eq:Higgs coupling matrix} in the light-neutrino mass basis with charged-lepton flavours, we project the light-neutrino components back onto the active-flavour basis using the \LO active-light mixing matrix \eqref{eq:light-neutrino mixing matrix},
\begin{align} \label{eq:projected neutral amplitudes}
 \mat Z^{\nu N} &\equiv \mat U^\nu \widehat{\mat Z}^{\nu N}, &
 \mat H^{\nu N} &\equiv (\mat U^\nu)^\ast
 \widehat{\mat H}^{\nu N} .
\end{align}
The corresponding \LO flavour weights of the \NC and Higgs interactions are
\begin{align} \label{eq:neutral weight vectors}
 \vec w_{0,a}^{\NC} &\equiv \weight(\mat Z_0^{\nu N} \vec e_a) , &
 \vec w_{0,a}^{NH} &\equiv \weight(\mat H_0^{\nu N} \vec e_a).
\end{align}
If non-unitarity is negligible, the \NC matrices simplify to
\begin{equation}
\mat Z_0^{\nu N} = \mat U_0^I \left[1 + \order*{\norm{\mat\eta^\nu}}\right] .
\end{equation}
Therefore, the flavour weights of the \NC interaction reduce to the flavour weights of the \CC interaction \eqref{eq:CC weight vector}.
Furthermore, neglecting terms suppressed by the light-neutrino mass in the Higgs interaction matrix \eqref{eq:projected neutral amplitudes}, it simplifies to
\begin{equation}
\mat H_0^{\nu N} = (\mat U_0^I)^\ast\mat D^N
\left[1 + \order*{\norm{\mat\eta^\nu}} + \order*{\frac{m_\nu}{m_N^{}}}\right] .
\end{equation}
Since complex conjugation and the heavy-mass matrix do not change the flavour weight, the flavour weight of the Higgs interaction also reduces to the flavour weight of the \CC interaction \eqref{eq:CC weight vector}.
Under these approximations, one weight vector describes the flavour structure of the interactions between the pseudo-Dirac pair and the $W$, $Z$, and Higgs bosons.

At \NLO the active-heavy interaction matrix \eqref{eq:light-heavy interaction matrix} consists of the \LO term \eqref{eq:LO light-heavy interaction matrix} and the \NLO contribution \eqref{eq:NLO light-heavy interaction matrix}.
Since their directions satisfy
\begin{align}
 \inner{\vec e_-^\trans\vec e_-^\ast}
 &=
 \inner{\vec e_+^\trans\vec e_+^\ast}
 =
 1,
 &
 \inner{\vec e_-^\trans\vec e_+^\ast}
 &=
 0,
\end{align}
the interference term in the pair-summed product vanishes:
\begin{align} \label{eq:NLO pair summed product}
 \mat U^I (\mat U^I)^\dagger
 &=
 \mat U_0^I (\mat U_0^I)^\dagger
 +
 \order{\epsilon^2}.
\end{align}
The weight map \eqref{eq:weight map} therefore remains unchanged at linear order in the \LN-violating parameters with respect to the \LO result \eqref{eq:CC weight vector}:
\begin{align} \label{eq:NLO pair summed weight}
 \weight(\mat U^I)
 &=
 \weight(\mat U_0^I)
 +
 \order{\epsilon^2}
 =
 \weight(\vec d_{\slashed L}^\theta)
 +
 \order{\epsilon^2} .
\end{align}
This cancellation is specific to the pair-summed weight.
For a single heavy mass eigenstate,
\begin{align}
 (\mat U_0^I+\mat U_1^I)\vec e_a
 &= \vec c_0^I (\vec e_-)_a + \vec c_1^I (\vec e_+)_a + \order{\epsilon^2},
\end{align}
the linear interference does not vanish.
Here the \LO and \NLO interaction coefficient vectors are defined in \eqref{eq:LO light-heavy interaction matrix,eq:NLO light-heavy interaction matrix}, respectively.

\subsection{Flavour weights and light-neutrino oscillations}

The weight vector \eqref{eq:LO CC weight vector} depends only on the neutrino masses \eqref{eq:light-neutrino masses}, the light-neutrino Takagi matrix \eqref{eq:light-neutrino Takagi matrix}, and the Takagi sign \eqref{eq:Takagi sign}.
After the oscillation parameters \eqref{eq:light-neutrino oscillation observables} and the mass ordering \eqref{eq:light-neutrino masses} have been fixed, the only remaining inputs to the flavour weight are the Majorana phase \eqref{eq:Majorana phase} and the Takagi sign.
Changing the Takagi sign is equivalent to shifting the Majorana phase,
\begin{align} \label{eq:Takagi sign phase shift}
 \sigma e^{i\alpha}
 =
 e^{i\left(\alpha+\frac{1-\sigma}{2}\pi\right)} .
\end{align}
We absorb this shift into the phase variable.
This can be directly compared to the treatment in the usual \CI parametrisation \cite{Molinaro:2009lud,Granelli:2022eru}.
The active flavour direction \eqref{eq:light-neutrino direction} becomes
\begin{align} \label{eq:BF CC vector majorana phase}
 \vec d_{\slashed L}^\theta
 &\equiv
 \frac{\sqrt{m_\two}\vec U_\two
 - i e^{i\alpha}
 \sqrt{m_\three}\vec U_\three}{\sqrt{m_\two+m_\three}}, &
 (\vec U_a)_\ell &\equiv (\mat U_{\PMNS})_{\ell a} ,
\end{align}
where the flavour index is $\ell=e$, $\mu$, $\tau$, and the massive light-neutrino index is $a=\two$, $\three$.
The corresponding flavour weight \eqref{eq:CC weight vector} traces an ellipse as a function of the Majorana phase,
\begin{align} \label{eq:weight vector}
 \vec w_0 &\equiv \vec a + \vec d , &
 \vec d &\equiv \vec b \sin\alpha + \vec c \cos\alpha .
\end{align}
The ellipse is centred on the mass-weighted flavour average
\begin{equation} \label{eq:BF centre}
 \vec a \equiv
 \frac{m_\two \main(\tensor{\vec U_\two \vec U_\two^\dagger})
 + m_\three \main(\tensor{\vec U_\three \vec U_\three^\dagger})}{m_\two+m_\three} ,
\end{equation}
while its displacement is determined by the mass-weighted flavour overlap
\begin{align} \label{eq:BF displacement}
 \vec b &\equiv \Re\vec z, &
 \vec c &\equiv \Im\vec z, &
 \vec z &\equiv 2 \frac{\sqrt{m_\two m_\three}}{m_\two+m_\three} \main(\tensor{\vec U_\three \vec U_\two^\dagger}) ,
\end{align}
where the $\main$ function is defined alongside the weight map \eqref{eq:weight map}.
Unitarity of the \PMNS matrix implies
\begin{align} \label{eq:BF ellipse simplex constraints}
 \sum_{\ell=e,\mu,\tau}
 (\vec a)_\ell &= 1,
 &
 \sum_{\ell=e,\mu,\tau}
 (\vec b)_\ell =
 \sum_{\ell=e,\mu,\tau}
 (\vec c)_\ell &= 0 .
\end{align}
The coefficients \eqref{eq:BF centre,eq:BF displacement} of the ellipse \eqref{eq:weight vector} can be expanded for small $s_{13} \approx 0.15$.
For \NO, the ellipse \eqref{eq:weight vector}, with parameters
\begin{subequations} \label{eq:BF NO analytic structure}
\begin{align}
 \vec a_{\NO}^{}
 &=
 \frac{1}{1+r_{\two\three}}
 \begin{pmatrix}
 r_{\two\three} s_{12}^2 \\
 s_{23}^2 + r_{\two\three} c_{12}^2 c_{23}^2 \\
 c_{23}^2 + r_{\two\three} c_{12}^2 s_{23}^2
 \end{pmatrix}
 + \order{s_{13}},
 &
 \vec c_{\NO}^{}
 &= \order{s_{13}},
 \\
 \vec b_{\NO}^{}
 &=
 \frac{2\sqrt{r_{\two\three}}}{1+r_{\two\three}}
 c_{12} c_{23} s_{23}
 \begin{pmatrix}
 0 \\
 1 \\
 -1
 \end{pmatrix}
 + \order{s_{13}} , &
 r_{\two\three} &\equiv \frac{m_\two}{m_\three} \ll 1 .
\end{align}
\end{subequations}
is therefore centred near a $\mu$-$\tau$-dominated flavour point and elongated along the $\mu$-$\tau$ direction.
For \IO, the ellipse \eqref{eq:weight vector}, with parameters
\begin{subequations} \label{eq:BF IO analytic structure}
\begin{align}
 \vec a_{\IO}^{}
 &=
 \frac{1}{1+r_{\two\three}}
 \begin{pmatrix}
 s_{12}^2 + r_{\two\three} c_{12}^2\\
 c_{12}^2 c_{23}^2 + r_{\two\three} s_{12}^2 c_{23}^2 \\
 c_{12}^2 s_{23}^2 + r_{\two\three} s_{12}^2 s_{23}^2
 \end{pmatrix}
 + \order{s_{13}},
 &
 \vec c_{\IO}^{}
 &= \order{s_{13}},
 \\
 \vec b_{\IO}^{}
 &=
 \frac{2\sqrt{r_{\two\three}}}{1+r_{\two\three}}
 s_{12} c_{12}
 \begin{pmatrix}
 1 \\
 -c_{23}^2 \\
 -s_{23}^2
 \end{pmatrix}
 + \order{s_{13}} , &
 r_{\two\three} &\equiv \frac{m_\two}{m_\three} \lessapprox 1 .
\end{align}
\end{subequations}
is centred near a more democratic flavour point and elongated from an $e$-dominated flavour point towards a $\mu$-$\tau$-dominated flavour point.
At \LO in $s_{13}$, the flavour weights are independent of the \CP-violating Dirac phase \eqref{eq:PMNS invariant}; therefore, the large uncertainty of its \BF value enters only through terms suppressed by $s_{13}$.

\begin{figure}
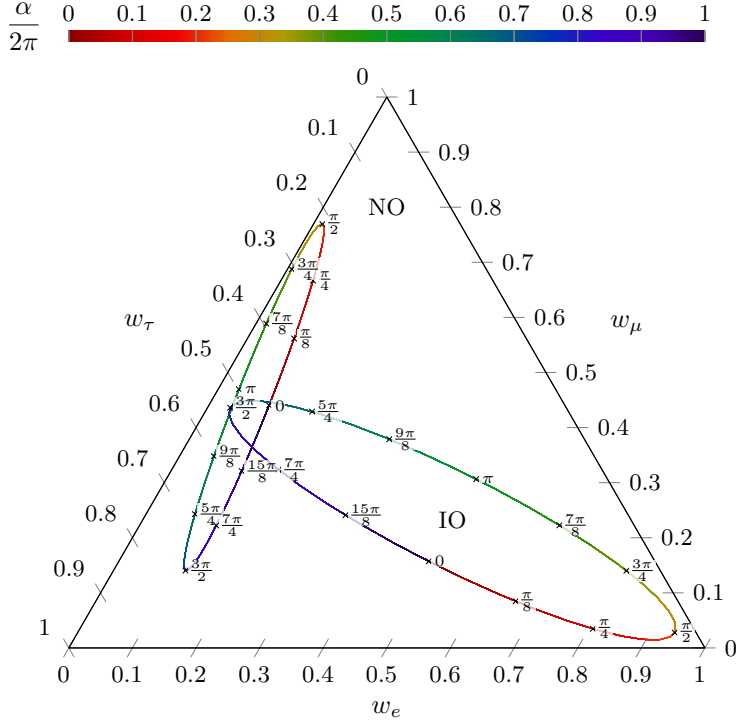

\begin{panels}{.65}
\includepgf{best-fit}
\end{panels}
\caption[\sentence\BFlong trajectories of the \LOlong flavour weights]{
$\BF$ trajectories of the active-heavy interaction flavour weights in the flavour simplex.
The free curve parameter is the Majorana phase \eqref{eq:Majorana phase}, and the two curves correspond to the \NO and \IO values in \eqref{eq:BF numerical curves}.
The labelled points mark selected values of the Majorana phase.
} \label{fig:best-fit}
\end{figure}

The light-neutrino oscillation parameters \eqref{eq:light-neutrino oscillation observables} are fitted by the \NuFIT collaboration to the available measurements, including observations from \IC \cite{IceCubeCollaboration:2024ssx} and \SK \cite{Super-Kamiokande:2023jbt}.
Using the \NuFIT~6.1 \cite{Esteban:2024eli} \IC{}24 \BF values of the oscillation inputs with \SK/\IC atmospheric data, the weight vector \eqref{eq:weight vector} becomes
\begin{subequations} \label{eq:BF numerical curves}
\begin{align}
 \vec w_{\NO}^{\BF}
 &\approx
 \begin{pmatrix*}[l]
 0.0637 \\
 0.455 \\
 0.481
 \end{pmatrix*}
 +
 \begin{pmatrix*}[l]
 -0.0496 \\
 \phantom{-}0.314 \\
 -0.265
 \end{pmatrix*}
 s_\alpha
 +
 \begin{pmatrix}
 \phantom{-}0.0310 \\
 -0.0146 \\
 -0.0164
 \end{pmatrix} c_\alpha,
 \\
 \vec w_{\IO}^{\BF}
 &\approx
 \begin{pmatrix}
 0.487 \\
 0.232 \\
 0.281
 \end{pmatrix}
 +
 \begin{pmatrix}
 \phantom{-}0.452 \\
 -0.204 \\
 -0.247
 \end{pmatrix}
 s_\alpha
 +
 \begin{pmatrix}
 0 \\
 -0.0746 \\
 \phantom{-}0.0746
 \end{pmatrix} c_\alpha,
\end{align}
\end{subequations}
where $c_\alpha\equiv \cos\alpha$ and $s_\alpha\equiv \sin\alpha$.
The corresponding trajectories in the flavour simplex are shown in \cref{fig:best-fit}.
The \NO curve lies close to the $\mu$-$\tau$ edge and follows mainly the $\mu$-$\tau$ direction.
By contrast, the \IO curve is centred at larger electron flavour and varies mainly between the electron corner and the opposite edge, as anticipated from the approximations in \eqref{eq:BF NO analytic structure,eq:BF IO analytic structure}.

\subsection{Uncertainties from light-neutrino oscillation measurements}

The fit to the light-neutrino oscillation measurements assigns a chi-square value $\chi^2(\vec x)$ to the observables \eqref{eq:light-neutrino oscillation observables}.
At the \BF point it takes the value $\chi_{\BF}^2$, and its relative increase is
\begin{equation} \label{eq:delta chi square}
 \Delta\chi^2(\vec x)
 \equiv
 \chi^2(\vec x)-\chi_{\BF}^2 .
\end{equation}
The corresponding \PDF for the oscillation inputs is
\begin{equation} \label{eq:oscillation likelihood}
 f(\vec x) \equiv
 \frac1{N}
 \exp\left[-\frac12\Delta\chi^2(\vec x)\right],
\end{equation}
where $N$ is the normalisation.
This \PDF reduces to a multivariate normal distribution if \eqref{eq:delta chi square} is quadratic.

For a fixed value of the Majorana phase \eqref{eq:Majorana phase}, the flavour weight \eqref{eq:weight vector} is a deterministic function of the oscillation inputs.
Its induced \PDF is the push-forward
\begin{equation} \label{eq:fixed phase push forward}
 f_\alpha(\vec w)
 \equiv
 \int
 f(\vec x)
 \delta^{(2)}(\vec w-\vec w(\vec x,\alpha))
 \d{\vec x}.
\end{equation}
If the Majorana phase is marginalised uniformly over its physical range, the corresponding \PDF is
\begin{equation} \label{eq:push forward}
 f(\vec w)
 \equiv
 \int_0^{2\pi}
 f_\alpha(\vec w)
 \differential\frac{\d{\alpha}}{2\pi} .
\end{equation}
The contour levels are given by highest-density regions of this \PDF.
For highest-density regions, it is convenient to use the complementary \CDF
\begin{equation} \label{eq:contour definition}
 F(t)
 \equiv
 \int
 f(\vec w)
 \Theta(f(\vec w)-t)
 \d^2\vec w ,
\end{equation}
where $t$ is a density threshold.
The threshold $t_Z^{}$ is quoted as a Gaussian-equivalent significance by matching the enclosed probability to the central probability of a one-dimensional normal distribution \cite{Cowan:2010js}
\begin{equation} \label{eq:sigma probabilities}
 F(t_Z^{})
 =
 \erf \frac{Z}{\sqrt2}.
\end{equation}

\begin{figure}
\begin{panels}{2}
\includepgf{majorana-scan-simple}
\caption{Uncorrelated likelihood} \label{fig:uncorrelated likelihood}
\panel
\includepgf{majorana-scan}
\caption{Correlated projected likelihood} \label{fig:correlated projected likelihood}
\end{panels}
\caption[Phase-marginalised likelihood regions for the \LOlong flavour weights]{
Phase-marginalised likelihood regions for the flavour weights of the \LO active-heavy interactions.
The contours correspond to highest-density regions of the flavour-weight \PDF \eqref{eq:push forward}, with the Majorana phase uniformly marginalised.
The uncorrelated likelihood is shown in panel \subref{fig:uncorrelated likelihood}, whereas the projected-likelihood reconstruction \eqref{eq:oscillation composite likelihood} is shown in panel \subref{fig:correlated projected likelihood}.
The overlaid \BF trajectories are those of \cref{fig:best-fit}.
} \label{fig:marginalised majorana likelihood}
\end{figure}

\begin{description}
\item[Uncorrelated likelihood]

The reported asymmetric one-dimensional uncertainties around the \BF values are used to build independent one-dimensional \PDFs for the six oscillation parameters \eqref{eq:light-neutrino oscillation observables}.
Their product defines an uncorrelated approximation to the \PDF \eqref{eq:oscillation likelihood}.
This approximation is transparent and gives the correct scale of the uncertainty, but it discards correlations between oscillation parameters.
The resulting likelihood regions \eqref{eq:push forward} are presented in \cref{fig:uncorrelated likelihood}.

\item[Correlated projected likelihood]

For a block of observables $\vec y$, the projected likelihood is obtained by profiling over the complementary observables $\bar{\vec y}$:
\begin{equation} \label{eq:projected likelihood definition}
 \Delta\chi_p^2(\vec y)
 \equiv
 \min_{\bar{\vec y}}
 \Delta\chi^2(\vec y, \bar{\vec y}),
\end{equation}
where the right-hand side is the full likelihood before projection, minimised over the observables not contained in $\vec y$.
Thus, for example, $\Delta\chi_p^2(a,b)$ denotes the two-dimensional projection onto the variables $a$ and $b$.
Such a projection is more informative than a one-dimensional uncertainty because it retains correlations inside the projected block.
The public \NuFIT release gives the atmospheric variables as a three-dimensional projection and the remaining information as one- and two-dimensional projections.
The solar-reactor observables are
\begin{equation} \label{eq:solar reactor projection variables}
 \vec y_S^{} \equiv
 \row{
 \Delta m_{21}^2,
 s_{12}^2,
 s_{13}^2
 }^\trans ,
\end{equation}
while the atmospheric observables are
\begin{align} \label{eq:atmospheric projection variables}
 \vec y_A^{} &\equiv
 \row{
 \Delta m_{3k}^2,
 s_{23}^2,
 \delta
 }^\trans , &
 k &\equiv \begin{cases}
 1 & \text{for } \NO \\
 2 & \text{for } \IO
 \end{cases}.
\end{align}
The pairwise composite approximation to the full oscillation likelihood is
\begin{equation} \label{eq:oscillation composite likelihood}
 \Delta\chi_{\text{proj}}^2(\vec x)
 \equiv
 \Delta\chi_p^2(\vec y_A^{})
 + \Delta\chi_p^2(\vec y_S^{})
 + \Delta\chi_{\text{cross}}^2(\vec y_A^{},\vec y_S^{}),
\end{equation}
where $\Delta\chi_p^2(\vec y_A^{})$ is the full three-dimensional atmospheric projection.
The three-variable solar-reactor contribution is reconstructed from the one- and two-dimensional projections by retaining all pairwise correlations:
\begin{align} \label{eq:oscillation composite likelihood 2}
 \Delta\chi_p^2(\vec y_S^{})
 &=
 \sum_{\substack{a,b\in\vec y_S^{}\\a<b}}
 \Delta\chi_p^2(a,b)
 -
 \sum_{a\in\vec y_S^{}}
 \Delta\chi_p^2(a).
\end{align}
Here $a<b$ refers to the order of entries in $\vec y_S^{}$.
The published pairwise correlations between atmospheric and solar-reactor variables are incorporated through
\begin{align}
 \Delta\chi_{\text{cross}}^2(\vec y_A^{},\vec y_S^{})
 &\equiv
 \sum_{a\in\vec y_A^{}}
 \sum_{b\in\vec y_S^{}}
 \left[
 \Delta\chi_p^2(a,b)
 - \Delta\chi_p^2(a)
 - \Delta\chi_p^2(b)
 \right].
\end{align}
The resulting density contours \eqref{eq:push forward} are presented in \cref{fig:correlated projected likelihood}.
\end{description}

\begin{figure}
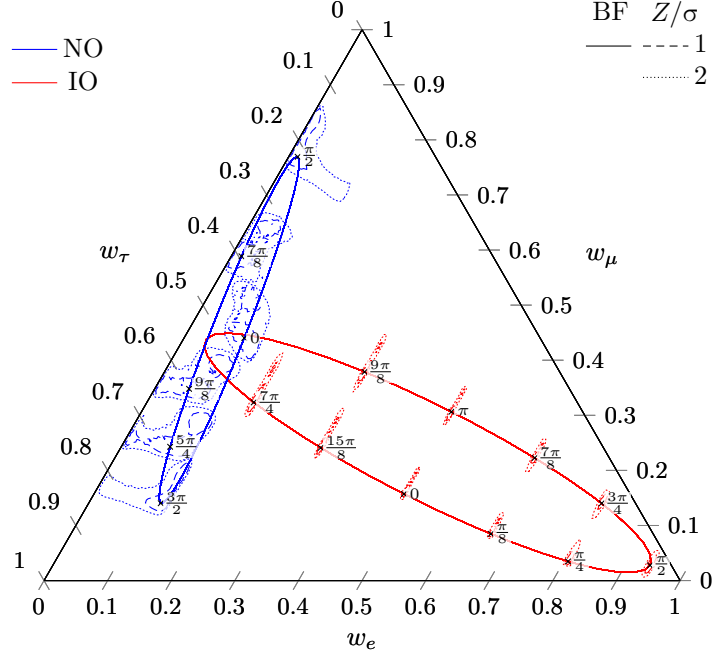

\begin{panels}{.65}
\includepgf{majorana-fixed}
\end{panels}
\caption[Fixed-phase likelihood regions for the projected-likelihood reconstruction]{
Selected fixed-phase likelihood regions for the projected-likelihood reconstruction.
Each group of contours is obtained from the fixed-phase \PDF \eqref{eq:fixed phase push forward} for one selected value of the Majorana phase.
The corresponding \BF points on the \NO and \IO trajectories are indicated by labelled markers.
The phase-marginalised regions in \cref{fig:correlated projected likelihood} are recovered by sweeping the fixed-phase oscillation uncertainty over the Majorana phase.
The uncertainty regions grow with a reduced electron content.
} \label{fig:fixed majorana likelihood}
\end{figure}

The fixed-phase construction is shown in \cref{fig:fixed majorana likelihood}.
For a fixed Majorana phase, the oscillation likelihood produces a two-dimensional region around the corresponding \BF point.
Integrating these fixed-phase densities over the Majorana phase gives the phase-marginalised densities in \cref{fig:correlated projected likelihood}.

\subsection{\sentence\NNOlong} \label{sec:heavy-neutrino oscillations}

The coherent propagation of the pseudo-Dirac pair can produce \NNOs between \LN-conserving and -violating decays \cite{Antusch:2020pnn}.
For the present purpose it is sufficient to use the simplified \NNO formula: wave-packet suppression, localisation effects, and spin correlations are assumed not to modify the first oscillation periods \cite{Beuthe:2001rc, Antusch:2023nqd}.
The coherent time evolution in the heavy subspace is described in terms of the proper time $\tau$ by the heavy-neutrino evolution matrix.
The pseudo-Dirac mass splitting in \eqref{eq:heavy mass splitting} gives the relative oscillation phase
\begin{align} \label{eq:heavy propagation matrix}
 \mat E^N(\tau)
 &\equiv
 \exp\left(-i \mat D^N \tau\right), &
 \varphi_N(\tau)
 &\equiv
 \Delta m\tau
 +
 \order{\epsilon^2}.
\end{align}
The \LN-conserving amplitude $\mat A^+(\tau)$ leading to an \OS signature and the \LN-violating amplitude $\mat A^-(\tau)$ leading to a \SS signature are
\begin{align} \label{eq:heavy oscillation amplitudes}
 \mat A^+(\tau)
 &\propto
 e^{i m_N^{}\tau}
 \mat U^I
 \mat E^N(\tau)
 (\mat U^I)^\dagger,
 &
 \mat A^-(\tau)
 &\propto
 e^{i m_N^{}\tau}
 (\mat U^I)^\ast
 \mat E^N(\tau)
 (\mat U^I)^\dagger.
\end{align}
In probability calculations, one either selects a specific production flavour or sums over all of them.
Moreover, the \LN information is retained only if the two \LN processes are treated together.
Therefore, we define the combined amplitudes
\begin{align} \label{eq:combined amplitude}
 \mat A(\tau) &=
\begin{pmatrix}
 \mat A^+(\tau) \\
 \mat A^-(\tau)
 \end{pmatrix} , &
 \vec A_\ell(\tau)
 &\equiv
 \begin{pmatrix}
 \vec A_\ell^+(\tau) \\
 \vec A_\ell^-(\tau)
 \end{pmatrix} , &
 \vec A_\ell^\pm(\tau) &\equiv \mat A^\pm(\tau) \vec e_\ell ,
\end{align}
where $\vec e_\ell$ is a flavour basis vector selecting the flavour $\ell$.
The globally normalised flavour-probability matrix for \LN-conserving and -violating processes is
\footnote{
The Hadamard product $(\mat A \odot \mat B)_{ij} = (\mat A)_{ij} (\mat B)_{ij}$ corresponds to element-wise multiplication.
}
\begin{align} \label{eq:flavour probability matrix}
 \mat P^\pm(\tau)
 &\equiv
 \frac{
 (\mat A^\pm(\tau))^\ast \odot \mat A^\pm(\tau)
 }{
 \norm*{\mat A(\tau)}^2
 }
 .
\end{align}
The production-summed decay-flavour probability vector is obtained from this matrix by summing over the production flavour.
For a fixed production flavour $\ell$, the corresponding conditional vector is instead normalised with the fixed-production amplitude.
\begin{align} \label{eq:flavour probability vector}
 \vec P^\pm(\tau) &\equiv \mat P^\pm(\tau) \vec 1
 , &
 \vec P_\ell^\pm(\tau) &\equiv
 \frac{
  \mat P^\pm(\tau) \vec e_\ell
 }{
 \inner{
 \vec 1^\trans
 \left[
 \mat P^+(\tau)+\mat P^-(\tau)
 \right]
 \vec e_\ell
 }
 }
 .
\end{align}
The probability vectors are obtained directly by applying the weight map \eqref{eq:weight map} to the combined amplitudes \eqref{eq:combined amplitude}.
\begin{align}
 \begin{pmatrix}
 \vec P^+(\tau) \\
 \vec P^-(\tau)
 \end{pmatrix}
 &=
 \weight(\mat A(\tau))
 , &
 \begin{pmatrix}
 \vec P_\ell^+(\tau)\\
 \vec P_\ell^-(\tau)
 \end{pmatrix}
 &=
 \weight(\vec A_\ell(\tau))
 .
\end{align}
Summing over all decay flavours gives the flavour-independent \LN probabilities
\begin{align} \label{eq:LN probability}
 P^\pm(\tau)
 &\equiv
 \frac{\norm*{\mat A^\pm(\tau)}^2}
 {\norm*{\mat A(\tau)}^2} =
 \inner{\vec 1^\trans \vec P^\pm(\tau)}
 , &
 P_\ell^\pm(\tau) &\equiv \frac{
 \abs{\vec A_\ell^\pm(\tau)}^2
 }{
 \abs{\vec A_\ell(\tau)}^2
 } = \inner{\vec 1^\trans \vec P_\ell^\pm(\tau)}
 .
\end{align}
They give the probabilities that an event belongs to the \LN-conserving or -violating channel.
\begin{align}
 P^+(\tau) + P^-(\tau)
 =
 P_\ell^+(\tau) + P_\ell^-(\tau)
 &=
 1.
\end{align}
The corresponding flavour weight vectors are
\begin{align} \label{eq:probability weight}
 \vec w^\pm(\tau) &\equiv \weight(\mat A^\pm(\tau)) = \frac{\vec P^\pm(\tau)}{P^\pm(\tau)} , &
 \vec w_\ell^\pm(\tau) &\equiv \weight(\vec A_\ell^\pm(\tau)) = \frac{\vec P_\ell^\pm(\tau)}{P_\ell^\pm(\tau)}.
\end{align}

Substituting the \LO factorisation \eqref{eq:LO light-heavy interaction matrix} of the active-heavy interaction matrix \eqref{eq:light-heavy interaction matrix} into the amplitudes \eqref{eq:heavy oscillation amplitudes} leads to
\begin{align} \label{eq:heavy oscillation LO amplitudes}
 \mat A^+_0(\tau)
 &\propto
 \tensor{\vec c_0^I(\vec c_0^I)^\dagger}
 \cos\frac{\Delta m\tau}{2},
 &
 \mat A^-_0(\tau)
 &\propto
 i
 \tensor{(\vec c_0^I)^\ast(\vec c_0^I)^\dagger}
 \sin\frac{\Delta m\tau}{2}.
\end{align}
The resulting flavour-probability matrix \eqref{eq:flavour probability matrix} is
\begin{align} \label{eq:HNL joint probability LO}
 \mat P_0^\pm(\tau)
 &=
 \tensor{\vec w_0\vec w_0^\trans} c_0^\pm(\tau) , &
 c^\pm_0(\tau) &= \frac{1\pm\cos\Delta m\tau}{2}.
\end{align}
For the probability vector \eqref{eq:flavour probability vector}, the \LN probability \eqref{eq:LN probability}, and the weight vector \eqref{eq:probability weight}, fixing an initial production flavour or summing over all production flavours gives the same result.
\begin{align} \label{eq:heavy oscillation LO probabilities}
 \vec P_{\ell,0}^\pm(\tau)
 =
 \vec P_0^\pm(\tau)
 &= c^\pm_0(\tau) \vec w_0 , &
 P_{\ell,0}^\pm(\tau)= P_0^\pm(\tau)
 &=
 c_0^\pm(\tau), &
 \vec w_{\ell,0}^\pm(\tau) =
 \vec w_0^\pm(\tau)
 &=
 \vec w_0 .
\end{align}
Thus the leading \NNO exchanges the neutrino and antineutrino components of the pseudo-Dirac pair, while the charged-lepton flavour distribution remains the fixed vector \eqref{eq:CC weight vector}.

The \NLO active-heavy interaction matrix \eqref{eq:NLO light-heavy interaction matrix} gives the corresponding contributions to the amplitudes,
\begin{align} \label{eq:heavy oscillation amplitudes NLO}
 \mat A^+_1(\tau)
 &\propto
 i
 \left[
 \tensor{\vec c_1^I(\vec c_0^I)^\dagger}
 +
 \tensor{\vec c_0^I(\vec c_1^I)^\dagger}
 \right]
 \sin\frac{\Delta m\tau}{2},
 &
 \mat A^-_1(\tau)
 &\propto
 \left[
 \tensor{(\vec c_1^I)^\ast(\vec c_0^I)^\dagger}
 +
 \tensor{(\vec c_0^I)^\ast(\vec c_1^I)^\dagger}
 \right]
 \cos\frac{\Delta m\tau}{2}.
\end{align}
The interference between the \LO and \NLO amplitudes gives the \NLO contribution to the flavour probability matrix \eqref{eq:flavour probability matrix},
\begin{align} \label{eq:HNL joint probability NLO}
 \mat P_1^\pm(\tau)
 &=
 \left[
 2 c_0^\pm(\tau) w_1 \tensor{\vec w_0\vec w_0^\trans}
 - \tensor{\vec w_1\vec w_0^\trans}
 \pm \tensor{\vec w_0\vec w_1^\trans}
 \right]
 \sin\Delta m\tau ,
\end{align}
where the unnormalised interference vector and its component sum are
\begin{subequations} \label{eq:heavy oscillation CP invariant}
\begin{align}
 \vec w_1
 \equiv
 \frac{\Im\main \tensor{\vec c_1^I(\vec c_0^I)^\dagger}}{\abs{\vec c_0^I}^2}
 &=
 \frac{\Delta m_{\two\three}}{2\abs{\vartheta}^2m_N^{}} \vec I_{\nu N},
 &
 w_1
 \equiv
 \inner{\vec 1^\trans \vec w_1}
 =
 \frac{\Im \inner{(\vec c_1^I)^\trans(\vec c_0^I)^\ast}}{\abs{\vec c_0^I}^2}
 &=
 \frac{\Delta m_{\two\three}}{2\abs{\vartheta}^2m_N^{}}
 I_{\nu N}.
\end{align}
\end{subequations}
The dimensionless interference vector is
\begin{align}
 \vec I_{\nu N}
 &=
 (1-s) I_{\nu N} \vec w_0
 + \frac{2 s\ev{m_{\two\three}}}{\Delta m_{\two\three}} \vec w_{\nu N} , &
 \inner{\vec 1^\trans \vec I_{\nu N}}
 &=
 I_{\nu N}.
\end{align}
The explicit form of the interference follows from the coefficient vectors in \eqref{eq:LO light-heavy interaction matrix,eq:NLO light-heavy interaction matrix}.
The vector $\vec w_{\nu N}$ collects the phase-sensitive flavour direction of the \NLO oscillation,
\begin{align} \label{eq:NLO weight}
 \vec w_{\nu N}
 &\equiv
 I_{\nu N} \vec a^\prime
 -
 R_{\nu N} \vec d'.
\end{align}
The corresponding flavour directions are
\begin{align}
 \vec a^\prime
 &\equiv
 \frac{
 m_\three\main\tensor{\vec U_\three\vec U_\three^\dagger}
 -
 m_\two\main\tensor{\vec U_\two\vec U_\two^\dagger}
 }{m_\two+m_\three},
 &
 \vec d'
 &\equiv
 \frac{\partial\vec d}{\partial\alpha}
 =
 \vec b\cos\alpha
 -
 \vec c\sin\alpha .
\end{align}
The vector $\vec a'$ shifts the ellipse centre by the signed mass-weighted flavour imbalance, while $\vec d'$ generates motion along the leading flavour ellipse.

The corresponding \NLO decay-flavour probability vectors \eqref{eq:flavour probability vector} for fixed and summed production flavour are
\begin{align}
\label{eq:heavy oscillating charge resolved flavour fraction}
 \vec P^\pm_1(\tau)
 &= \left[
\pm w_1 \vec w_0
 -\Delta\vec w_1^\pm(\tau)
 \right] \sin\Delta m\tau
 , &
 \vec P_{\ell,1}^\pm(\tau)
 &=
 \left[
 \pm \frac{(\vec w_1)_\ell}{(\vec w_0)_\ell} \vec w_0
 -\Delta\vec w_1^\pm(\tau)
 \right]
 \sin\Delta m\tau
 .
\end{align}
The interference combination entering these probability vectors is
\begin{align} \label{eq:heavy oscillation centred CP invariant}
 \Delta\vec w_1^\pm(\tau) &= \vec w_1 - 2 w_1 \vec w_0 c_0^\pm(\tau) .
\end{align}
The corrections to the scalar \LN probabilities \eqref{eq:LN probability} are
\begin{align}
 P_1^\pm(\tau)
 &=
 \pm
 \left[
 w_1 + w_1 \cos \Delta m \tau
 \right]
 \sin\Delta m\tau
, &
 P_{\ell,1}^\pm(\tau)
 &=
 \pm
 \left[
 \frac{(\vec w_1)_\ell}{(\vec w_0)_\ell}
 +
 w_1 \cos\Delta m\tau
 \right]
 \sin\Delta m\tau
 .
\end{align}
The corrections to the flavour weights \eqref{eq:probability weight} are
\begin{align} \label{eq:heavy oscillating centred flavour fraction}
 \vec w_{\ell,1}^\pm(\tau)
 = \vec w_1^\pm(\tau)
 &=
 \frac{
 \vec P_1^\pm(\tau)
 -
 P_1^\pm(\tau)\vec w_0
 }{c_0^\pm(\tau)}
 =
 -\frac{\Delta\vec w_1}{c_0^\pm(\tau)}
 \sin\Delta m\tau,
 &
 \Delta\vec w_1
 &\equiv
 \vec w_1
 -
 w_1
 \vec w_0 .
\end{align}
Therefore, at \NLO the leading \LN oscillation receives a correction controlled by $w_1$, while a genuine flavour oscillation with amplitude proportional to $\Delta\vec w_1$ appears.

The \NLO flavour-weight correction discussed in \cref{sec:active-heavy flavour weights} vanishes for an incoherent pair-summed interaction because the heavy directions $\vec e_-$ and $\vec e_+$ are orthogonal.
It reappears in oscillations because coherent time evolution inserts the non-trivial propagation matrix \eqref{eq:heavy propagation matrix}, turning the same \LO-\NLO structure into a time-dependent interference effect.

\subsection{\sentence\NDBlong decay} \label{sec:0nubb}

\resetacronym{NDB}

\begin{figure}
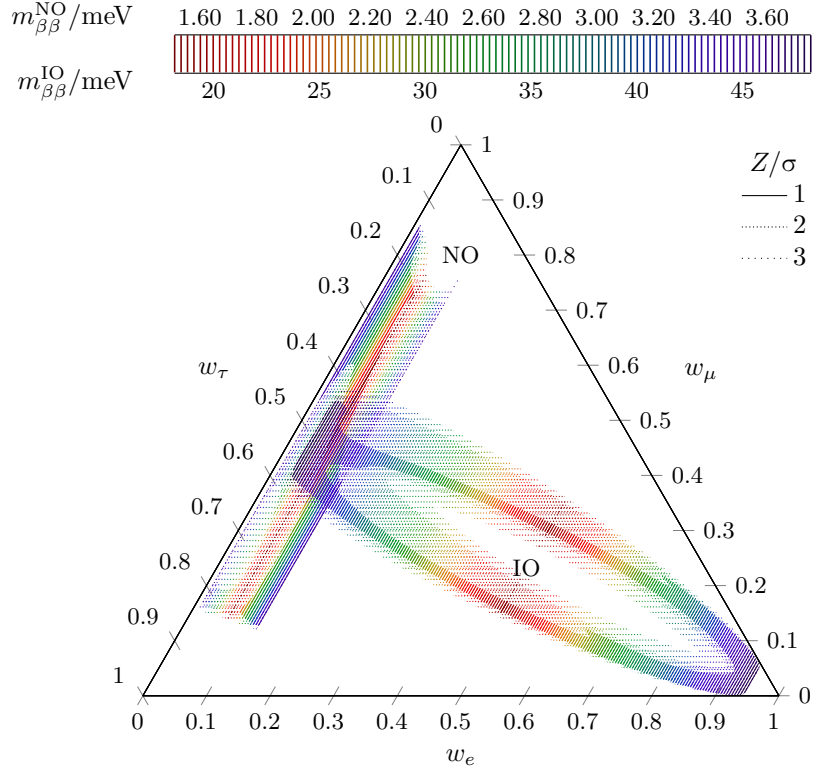

\begin{panels}{.65}
\includepgf{neutrinolessdoublebeta}
\end{panels}
\caption[Light-neutrino \NDBlong effective mass]{
Light-neutrino $\NDB$ effective-mass values on the phase-marginalised flavour-weight likelihood regions.
The likelihood contours are taken from the projected-likelihood reconstruction in \cref{fig:correlated projected likelihood}.
} \label{fig:0nubb ternary likelihood}
\end{figure}

The \CC interaction \eqref{eq:charged current mass basis} also determines the neutrino-exchange contribution to \NDB decay \cite{Schechter:1980gr,Atre:2009rg,Deppisch:2012nb,DellOro:2016tmg}.
In the convention used in \eqref{eq:CC matrix}, the electron row of the full \CC matrix gives the \LN-violating exchange amplitude
\begin{equation} \label{eq:0nubb general amplitude}
 A_{\NDB}
 \propto
 \sum_{k=1}^5
 m_k \widehat{\mat W}_{e k}^2
 G(m_k) =
 \sum_{i=1}^3
 m_i (\mat U^\nu)_{e i}^2
 G(m_i)
 +
 \sum_{a=1}^2
 m_{3+a} (\mat U^I)_{e a}^2
 G(m_{3+a}) .
\end{equation}
Here $G(m)$ denotes the mass-dependent neutrino propagator, including the nuclear matrix-element factor \cite{Blennow:2010th,Deppisch:2012nb}.
For light neutrinos, $m_i$ is much smaller than the virtual momentum scale $p$ of the process, so that
\begin{equation} \label{eq:light propagator}
 G(m_i) = G_\nu
 \left[1+\order*{\frac{m_i^2}{p^2}}\right] .
\end{equation}
The coefficient $G_\nu$ is common to all light-neutrino mass eigenstates.
It therefore factorises from the light-neutrino amplitude in \eqref{eq:0nubb general amplitude} and is absorbed into the definition of the complex effective mass
\begin{equation} \label{eq:0nubb light effective mass}
 m_{\beta\beta}^\nu
 \equiv
 \sum_{i=1}^3
 m_i(\mat U^\nu)_{e i}^2
 =
 \sum_{i=1}^3
 m_i(\mat U_0^\nu)_{e i}^2
 +
 \order{\epsilon^2}.
\end{equation}
The light-neutrino contribution to the \NDB decay rate is controlled by its modulus.
At \LO in the \LN-violating parameters it reads
\begin{equation}
 m_{\beta\beta}^\nu
 = \sum_{i=1}^3 m_i
 \left[
 (\mat U_{\slashed L}^\nu)_{ei}^2
 -\frac{2(\vec \theta)_e^\ast}{s+s^2}
 (\mat U_{\slashed L}^\nu)_{ei}
 (\vec \theta^\trans\mat U_{\slashed L}^\nu)_i
 +
 \frac{[(\vec \theta)_e^\ast]^2}{(s+s^2)^2}
 (\vec \theta^\trans\mat U_{\slashed L}^\nu)_i^2
 \right]
 +
 \order{\epsilon^2}
 .
\end{equation}
If the non-unitarity of the active-light mixing matrix \eqref{eq:light-neutrino mixing matrix} is neglected, only the light-neutrino Takagi matrix \eqref{eq:Majorana phase} remains.
In terms of the centre and displacement of the flavour weight vector \eqref{eq:weight vector}, the modulus can then be expressed as
\begin{align} \label{eq:0nubb light}
 \abs{m_{\beta\beta}^\nu}
 =
 \abs{m_\two+m_\three}
 \sqrt{(\vec a)_e^2 - (\vec d)_e^2}
 =\begin{cases}
 \abs{
 m_2 s_{12}^2 c_{13}^2
 +
 m_3 s_{13}^2 e^{2i(\alpha-\delta)}
 } & \text{for } \NO, \\
 c_{13}^2
 \abs{
 m_1 c_{12}^2
 +
 m_2 s_{12}^2 e^{2i\alpha}
 } & \text{for } \IO .
 \end{cases},
\end{align}
where the explicit expression is given in the \PMNS convention \eqref{eq:PMNS matrix,eq:Majorana phase}; see also \cite{Bezrukov:2005mx,Drewes:2022akb}.
Its contour representation is shown in \cref{fig:0nubb ternary likelihood}.

\section{Conclusion} \label{sec:conclusion}

We have derived a perturbative Takagi description of the neutrino mixing matrix for a single pseudo-Dirac \HNL pair in the \SPSS.
The exact \LN-conserving Dirac limit is diagonalised without expanding in the active-sterile mixing, while the observed light-neutrino masses are introduced through a controlled expansion in the small \LN-violating parameters.
This gives analytic expressions for the light- and heavy-neutrino masses, the full block structure of the mixing matrix, and the active-light unitary defect.
The resulting parameter count shows that, once the light-neutrino masses and mixing parameters are fixed, the remaining continuous light-heavy information is a single complex amplitude.

The main new result is the symmetry-protected flavour reconstruction.
It reconstructs the active-heavy interaction matrix from the rank-two light-neutrino mass matrix and one complex light-heavy amplitude.
The modulus of this amplitude fixes the overall size of active-sterile mixing, while its phase defines a \CP-odd light-heavy invariant.
Thus the construction separates the flavour information fixed by the light-neutrino oscillations from the residual high-energy information.
In this sense, it implements the minimal-flavour idea directly in the pseudo-Dirac \SPSS, but at the level of the perturbative mixing matrix rather than through the conventional seesaw expansion.

A direct phenomenological consequence is obtained for normalised active-heavy interaction flavours.
For the pair-summed \CC interaction, the heavy-sector rotation and the light-heavy amplitude cancel from the leading flavour weight.
After neglecting active-light non-unitarity and terms suppressed by the smallness of the light-neutrino masses, the same vector also controls the leading light-heavy parts of the $Z$- and Higgs-boson interactions.
The leading flavour prediction is therefore fixed by the measured light-neutrino masses, the \PMNS matrix, and the Majorana phase.
It traces an ellipse in the flavour simplex, and the mapping of the \NuFIT uncertainties onto this prediction is presented in the ternary plots in \cref{fig:best-fit,fig:marginalised majorana likelihood,fig:fixed majorana likelihood}.

The linear \LN-violating corrections sharpen the distinction between incoherent and coherent observables.
For incoherent pair-summed flavour weights the \NLO interference cancels, so the leading flavour ellipse is stable at linear order.
For coherent heavy-neutrino propagation, the propagation matrix converts the same \LO-\NLO structure into a time-dependent flavour interference.
The heavy-neutrino oscillation result therefore contains a genuine \NLO flavour modulation controlled by the pseudo-Dirac mass splitting, the two active flavour directions induced by the light sector, and the light-heavy \CP invariant.

The same reconstruction also organises the \NDB and effective-operator results.
The light-neutrino contribution to \NDB decay can be written in terms of the same ellipse parameters that determine the leading flavour weights.
The heavy contribution separates into a pseudo-Dirac mass-splitting term and an interference term involving the \NLO interaction block.
After the pseudo-Dirac pair is integrated out, the coefficient of the Weinberg operator fixes the normalised flavour direction of the leading dimension-six operator, while the overall coefficient is controlled by the size of the light-heavy amplitude.
These relations make explicit which observables are fixed by light-neutrino data and which require the additional physical information encoded in the complex amplitude.

\subsection*{Acknowledgements}

This work is supported by the \FCT project \textnumero\ \href{https://doi.org/10.54499/2020.03969.CEECIND/CP1587/CT0014}{2020.\allowbreak 03969.\allowbreak CEECIND/\allowbreak CP1587/\allowbreak CT0014}.
The author thanks the Particles and Cosmology group at the University of Basel for their hospitality.

\appendix

\section{Comparison with the \CIlong construction} \label{sec:CI comparison}

\resetacronym{CI}

The parametrisation derived in \cref{sec:light-heavy interaction matrix} has a similar goal to the conventional \CI construction \cite{Casas:2001sr}, whose residual parameters and generalisations have been studied \eg in \cite{Xing:2009vb,Herrero-Garcia:2025aox}.
Both descriptions express active-heavy interaction structures in terms of light-neutrino data and residual high-energy information.
They differ, however, in the constraint that is solved, the regime of validity, and in the physical meaning of the residual freedom.

To derive the \CI parametrisation of the conventional \TIS, one first chooses a basis in which the heavy Majorana mass matrix is diagonal and positive.
At \LO in the seesaw expansion \eqref{eq:seesaw expansion}, the active-sterile mixing matrix, the light-neutrino mass matrix, and its Takagi diagonalisation are
\begin{align} \label{eq:seesaw relation}
 \mat \theta &\equiv \mat m_D^{}(\mat D^N)^{-1} , &
 \mat M_{\TIS}^\nu &\approx - \mat m_D^{} (\mat D^N)^{-1} \mat m_D^\trans , &
 \mat D^\nu &\equiv (\mat U_{\TIS}^\nu)^\trans\mat M_{\TIS}^\nu\mat U_{\TIS}^\nu .
\end{align}
The \CI solution of this quadratic constraint is
\begin{align} \label{eq:CI solution}
 \mat m_D^{} &= i(\mat U_{\TIS}^\nu)^\ast(\mat D^\nu)^{\nicefrac12}\mat R_{\CI} (\mat D^N)^{\nicefrac12} , &
 \mat R_{\CI}^\trans\mat R_{\CI} &= \mat 1.
\end{align}
Here $\mat R_{\CI}$ is a residual complex orthogonal matrix.
For two heavy neutrinos and one massless light neutrino, its light-neutrino rows can be ordered according to \eqref{eq:ordering dependent light indices}.
The residual matrix can then be written as
\begin{align} \label{eq:two heavy R}
 \mat R_{\CI}
 &\equiv
 \mat O_{\CI}\mat R(z),
 &
 \mat O_{\CI}
 &\equiv
\begin{psmallmatrix}
 0 & 0 \\
 1 & 0 \\
 0 & \xi
 \end{psmallmatrix} , &
 \xi &\in\{\pm1\}, &
 \mat R(z)
 &\equiv
 \begin{psmallmatrix}
 c_z & s_z \\
 -s_z & c_z
 \end{psmallmatrix}, &
 z &\equiv x+i y
 .
\end{align}
Here $c_z\equiv \cos z$ and $s_z\equiv \sin z$.
The sign labels the two disconnected components of the complex orthogonal square root of the seesaw constraint \eqref{eq:seesaw relation}.
Its origin is therefore the orthogonality condition \eqref{eq:CI solution}.
The absolute value of $y$ controls the amount of cancellation between different \LN-violating contributions to the model.
In the quasi-degenerate limit, this enhancement can be measured by
\begin{align}
\cosh(2y) &= \frac{1}{r_{\TIS}^{}} , &
r_{\TIS}^{} &= \frac{m_\two+m_\three}{m_N^{} \abs{\vartheta}^2}
\to \begin{cases}
1 & \text{canonical regime}, \\
0 & \text{cancellation regime},
\end{cases}
\end{align}
where $m_\two$ and $m_\three$ are the two non-zero light-neutrino masses \eqref{eq:light-neutrino masses}.

To make the flavour weight \eqref{eq:weight map} in this parameterisation transparent, we split the \CI solution \eqref{eq:CI solution} by introducing the \CI flavour amplitude
\begin{align} \label{eq:CI flavour amplitude}
 \mat m_D^{} &= i \mat A_{\CI} \mat R(x) (\mat D^N)^{\nicefrac12} , &
 \mat A_{\CI}
 &\equiv
 (\mat U_{\TIS}^\nu)^\ast
 (\mat D^\nu)^{\nicefrac12}
 \mat O_{\CI}
 \mat R(i y), &
 \mat R(z)
 &=
 \mat R(i y)
 \mat R(x).
\end{align}
In the quasi-degenerate heavy-pair limit, $\mat D^N\approx m_N^{}\mat 1$, and using the Majorana-phase convention of \eqref{eq:BF CC vector majorana phase}, the pair-summed flavour weight is determined only by the \CI flavour amplitude.
The reason is that the flavour map \eqref{eq:weight map} is invariant under multiplication by a non-zero scalar and right-multiplication by a unitary matrix.
\begin{align} \label{eq:CI pair summed weight}
 \vec w_{\CI}(\alpha,z,\xi)
 &\equiv
 \weight(\mat \theta)
 =
 \weight(\mat A_{\CI})
 =
 \vec a + r_{\CI}^{} \vec d .
\end{align}
This gives a family of concentric ellipses with the same orientation as \eqref{eq:weight vector}, but with the displacement \eqref{eq:BF displacement} rescaled by
\begin{align} \label{eq:CI scaling}
r_{\CI}^{} &= - \xi \tanh(2y)
=
- \xi \sgn(y)\sqrt{1- r_{\TIS}^2}
\to \begin{cases}
0 & \text{canonical regime}, \\
-\xi\sgn y & \text{cancellation regime}.
\end{cases}
\end{align}
For the canonical seesaw the ellipse shrinks to its centre \eqref{eq:BF centre}, while finite values of $y$ give concentric ellipses with the same orientation.
Only in the large-cancellation limit does the dependence on $y$ reduce to a sign.
For either sign in this limit, the resulting \CI weight coincides with the \SPSS ellipse \eqref{eq:weight vector} after, if necessary, shifting $\alpha\to\alpha+\pi$.

In the \SPSS construction of \cref{sec:light-heavy interaction matrix}, the starting point is the perturbative Takagi diagonalisation around the exact \LN-conserving Dirac limit.
After rotating to the light-neutrino mass basis, the constraint is the linear \LN-violating mass relation \eqref{eq:light-neutrino mass basis constraint}.
The residual freedom is the opposite rescaling of the two vectors \eqref{eq:rescaling parameterisation} that form the symmetric product.
It is described by the single complex light-heavy amplitude \eqref{eq:light-heavy amplitude}.
Its modulus fixes the size of active-sterile mixing \eqref{eq:active-sterile mixing}, while its phase is given by the \CP-odd light-heavy invariant \eqref{eq:light-heavy invariant}.
The Takagi sign appears in the Takagi vector \eqref{eq:Takagi sign} before any comparison with a heavy-sector square root is made.
It fixes the relative phase between the two massive light-neutrino directions so that the singular Takagi values are positive.

The two descriptions therefore have complementary validity ranges.
The standard \CI construction is tied to the seesaw relation \eqref{eq:seesaw relation}.
It is appropriate when the seesaw expansion \eqref{eq:seesaw expansion} is reliable and when higher-order non-unitarity corrections to the mixing matrix can be neglected.
It does not assume approximate \LN conservation and can be applied to a generic heavy Majorana spectrum.

The symmetry-protected flavour construction used in this paper instead assumes small \LN violation \eqref{eq:small LN violation} and a single accessible pseudo-Dirac pair.
Within that regime, the zeroth-order diagonalisation is exact in the active-sterile mixing \eqref{eq:active-sterile mixing}, and the expansion is performed in the small \LN-violating parameters.

Thus the \CI construction is the natural parametrisation of the generic seesaw in the seesaw limit, whereas the present symmetry-protected flavour construction is adapted to the symmetry-protected pseudo-Dirac regime and keeps the physical origin of the residual parameter manifest.

The master parametrisation of \cite{Cordero-Carrion:2018xre,Cordero-Carrion:2019qtu} extends the \CI construction to arbitrary Majorana neutrino-mass models.
Its starting point is the master equation
\begin{align}
 \mat m_\nu
 &=
 f
 \left(
 \mat y_1^\trans
 \mat M
 \mat y_2
 +
 \mat y_2^\trans
 \mat M^\trans
 \mat y_1
 \right),
 \label{eq:master equation}
\end{align}
which encompasses the canonical \TIS, \ISS, \LSS, and many radiative neutrino-mass models as special cases.
The canonical \TIS is recovered from \eqref{eq:master equation} through the identifications
\begin{align}
 f &= -1 , &
 \mat y_1 &= \mat y_2 = \frac{\mat m_D^{}}{v} , &
 \mat M &= v^2 \mat M_N^{-1} .
\end{align}
The master equation therefore reduces to the standard \TIS relation, while the corresponding master parametrisation reduces to the \CI parametrisation.

\section{\sentence\EFTlong description} \label{sec:minimal-flavour-reconstruction}

After integrating out the pseudo-Dirac pair, the leading effective operators are the \LN-violating dimension-five Weinberg operator \cite{Weinberg:1979sa} and the \LN-conserving dimension-six derivative operator,
\begin{align} \label{eq:minimal flavour EFT Lagrangian}
 \mathcal L_{\slashed L}^{(5)}
 &\equiv
 -\frac12
 (\mat c_{\slashed L}^{(5)})_{\alpha\beta}
 (\widetilde H^\dagger L_\alpha)
 (\widetilde H^\dagger L_\beta)
 +\hc,
 &
 \mathcal L_L^{(6)}
 &\equiv
 (\mat c_L^{(6)})_{\alpha\beta}
 (\widetilde H^\dagger L_\alpha)^\dagger
 i\widebar\sigma^\mu\partial_\mu
 (\widetilde H^\dagger L_\beta) .
\end{align}
For the minimal-flavour reconstruction \cite{Gavela:2009cd}, only the tree-level dimension-six operator generated by singlet-neutrino exchange is needed.
In a non-redundant SMEFT basis this operator can be rewritten as a fixed combination of Warsaw-basis operators, but it contains a single independent flavour coefficient at this order.

After \EWSB, the dimension-five operator contributes to the Majorana mass term, while the dimension-six operator contributes to the light-neutrino kinetic term,
\begin{align} \label{eq:minimal flavour EWSB operators}
 \mathcal L_{\slashed L}^{(5)}
 &\supset
 -\frac12
 \vec \nu^\trans
 v^2\mat c_{\slashed L}^{(5)}
 \vec \nu
 +\hc,
 &
 \mathcal L_L^{(6)}
 &\supset
 v^2
 (\mat c_L^{(6)})_{\alpha\beta}
 \nu_\alpha^\dagger
 i\widebar\sigma^\mu\partial_\mu
 \nu_\beta .
\end{align}
At the order relevant for the matching, the effective coefficients are
\begin{align} \label{eq:minimal flavour dim5 matching}
 v^2 \mat c_{\slashed L}^{(5)} &= \widehat{\mat M}_{\slashed L}^\nu ,&
 v^2\mat c_L^{(6)}
 &=
 \tensor{\vec \theta^\ast\vec \theta^\trans}
 \left[
 1+\order{\epsilon}
 \right].
\end{align}
The first relation identifies the coefficient of the Weinberg operator with the light-neutrino mass matrix \eqref{eq:light-neutrino mass matrix}.
The coefficients can be written in the interaction basis as
\begin{align} \label{eq:minimal flavour effective operators}
 v^2 \mat c_{\slashed L}^{(5)}
 &=
 \ev{m_{\two\three}}
 \left[
 \tensor{(\vec d_{\slashed L}^\theta)^\ast(\vec d_{\slashed L}^\mu)^\dagger}
 +
 \tensor{(\vec d_{\slashed L}^\mu)^\ast(\vec d_{\slashed L}^\theta)^\dagger}
 \right],
 &
 v^2\mat c_L^{(6)}
 &=
 \abs{\vartheta}^2
 \tensor{\vec d_{\slashed L}^\theta(\vec d_{\slashed L}^\theta)^\dagger}
 \left[
 1+\order{\epsilon}
 \right].
\end{align}
Here the \SM Higgs \VEV is given in \eqref{eq:Dirac mass}, the active-sterile mixing in \eqref{eq:active-sterile mixing}, and the physical light-heavy amplitude in \eqref{eq:light-heavy amplitude}.
The Weinberg coefficient is a symmetric bilinear between the active flavour directions in \eqref{eq:light-neutrino direction}.
For a single pseudo-Dirac pair, the dimension-five operator fixes the Takagi data entering these two directions, while the dimension-six operator is the rank-one projector onto the $\theta$-direction up to the overall normalisation $\abs{\vartheta}^2$.
Consequently, the rank-two light-neutrino mass matrix \eqref{eq:light-neutrino mass matrix}, or equivalently the coefficient of the Weinberg operator, fixes the flavour direction of the dimension-six operator up to an overall normalisation, as first observed in the minimal-flavour seesaw construction \cite{Gavela:2009cd}.
The normalised flavour information contained in the leading dimension-six operator is given by the weight vector \eqref{eq:weight map}
\begin{equation} \label{eq:minimal flavour reconstructed weight}
 \weight(\mat c_L^{(6)})
 =
 \weight(\vec d_{\slashed L}^\theta)
 =
 \vec w_0,
\end{equation}
which corresponds to the ellipse \eqref{eq:weight vector}.

\section{Heavy contribution to the \NDBlong decay} \label{sec:NLO NDB decay}

\resetacronym{NDB}

In the normalisation of \cref{sec:0nubb}, the \HNL exchange contribution to the \NDB decay can be quoted as a correction to the complex effective mass \cite{Atre:2009rg,Blennow:2010th}:
\begin{align} \label{eq:0nubb HNL effective mass correction}
 m_{\beta\beta}^{\nu+N}
 &\equiv
 \abs{m_{\beta\beta}^\nu+\delta m_{\beta\beta}^N} , &
 \delta m_{\beta\beta}^N
 &\equiv
 \frac{1}{G_\nu}
 \sum_{a=1}^2
 m_{3+a} (\mat U^I)_{e a}^2
 G(m_{3+a}) .
\end{align}
For parameter points for which the light-neutrino amplitude does not vanish, the induced shift of the absolute effective mass is, to linear order in the \HNL contribution,
\begin{align} \label{eq:0nubb absolute shift}
 m_{\beta\beta}^{\nu+N}
 &=
 \abs{m_{\beta\beta}^\nu}+\Delta m_{\beta\beta}^N , &
 \Delta m_{\beta\beta}^N
 &\equiv
 \Re
 \frac{
 (m_{\beta\beta}^\nu)^\ast
 \delta m_{\beta\beta}^N
 }{\abs{m_{\beta\beta}^\nu}}
 +
 \order*{
 \frac{\abs{\delta m_{\beta\beta}^N}^2}{\abs{m_{\beta\beta}^\nu}}
 } .
\end{align}
The expansion of \eqref{eq:light-heavy interaction matrix} gives
\begin{align}
 (\mat U^I)_{e a}^2
 &=
 (\mat U_0^I)_{e a}^2
 +
 2(\mat U_0^I)_{e a}(\mat U_1^I)_{e a}
 +
 \order{\epsilon^2}.
\end{align}
The first term gives the pseudo-Dirac mass-splitting contribution generated by the \LO interaction coefficient vector in \eqref{eq:LO light-heavy interaction matrix},
\begin{align} \label{eq:0nubb HNL LO}
 \delta m_{\beta\beta,0}^N
 &=
 (\vec c_0^I)_e^2
 \frac{G_N^-}{G_\nu} , &
 G_N^-
 &\equiv
 \sum_{a=1}^2
 (\vec e_-)_a^2
 m_{3+a} G(m_{3+a})= \frac{m_4 G(m_4) - m_5 G(m_5)}{2}
 ,
\end{align}
where the \LO interaction coefficient vector is defined in \eqref{eq:LO light-heavy interaction matrix}.
The interference contribution can be written as
\begin{equation} \label{eq:0nubb HNL interference}
 \delta m_{\beta\beta,1}^N
 \equiv
 \frac{2}{G_\nu}
 \sum_{a=1}^2
 m_{3+a}
 (\mat U_0^I)_{e a}
 (\mat U_1^I)_{e a}
 G(m_{3+a})
 =
 \frac{2}{G_\nu}
 (\vec c_0^I)_e
 (\vec c_1^I)_e
 G_N^+,
\end{equation}
where the \NLO interaction coefficient vector is defined in \eqref{eq:NLO light-heavy interaction matrix} and
\begin{equation}
 G_N^+
 \equiv
 \sum_{a=1}^2
 (\vec e_-)_a
 (\vec e_+)_a
 m_{3+a}G(m_{3+a})
 =
 \frac{
 m_4 G(m_4)
 +
 m_5 G(m_5)
 }{2}.
\end{equation}
The \HNL contribution that is linear in the \LN-violating parameters is therefore
\begin{align} \label{eq:0nubb HNL linear correction}
 \delta m_{\beta\beta}^N
 &=
 \delta m_{\beta\beta,0}^N
 +
 \delta m_{\beta\beta,1}^N
 +
 \order{\epsilon^2}.
\end{align}
For a heavy neutrino with mass much larger than $p$, the mass-dependent factor and the derivative needed below have the asymptotic forms \cite{Blennow:2010th}
\begin{align} \label{eq:heavy propagator}
 G(m)
 &=
 \frac{G_N}{m^2}
 \left[
 1+\order*{\frac{p^2}{m^2}}
 \right] , &
 \eval*{\dv{m}\left[m G(m)\right]}_{m=m_N^{}}
 &=
 -\frac{G_N}{m_N^2}
 \left[
 1+\order*{\frac{p^2}{m_N^2}}
 \right] .
\end{align}
The phase convention in \eqref{eq:light-heavy phase convention} gives $(\vec e_-)_1^2=\flatfrac12$ and $(\vec e_-)_2^2=-\flatfrac12$.
The same convention gives $(\vec e_-)_a(\vec e_+)_a=\flatfrac12$ for both heavy states.
Expanding the pseudo-Dirac pair around the average mass using \eqref{eq:heavy mass splitting} gives
\begin{equation}
G_N^-
 =
 -\frac{\Delta m}{2}
 \eval*{\dv{m}\left[m G(m)\right]}_{m=m_N^{}}
 +
 \order{\epsilon^3}
 =
 G_N
 \frac{\Delta m}{2m_N^2}
 \left[
 1+\order*{\frac{p^2}{m_N^2}}
 \right]
 +
 \order{\epsilon^3},
\end{equation}
and
\begin{equation}
 G_N^+
 =
 m_N^{}G(m_N^{})
 +
 \order{\epsilon^2}
 =
 \frac{G_N}{m_N^{}}
 \left[
 1+\order*{\frac{p^2}{m_N^2}}
 \right]
 +
 \order{\epsilon^2}.
\end{equation}
Using \eqref{eq:heavy propagator}, the two \HNL contributions combine to
\begin{equation}
 \delta m_{\beta\beta}^N
 = \frac{G_N}{G_\nu m_N} \left(2 (\vec c^I_0)_e (\vec c^I_1)_e + \frac{\Delta m}{2m_N} (\vec c^I_0)_e^2 \right)
 \left[
 1+\order*{\frac{p^2}{m_N^2}}
 \right].
\end{equation}
Consequently, the light-neutrino contribution \eqref{eq:0nubb light} is fixed by the light-sector parameters, whereas the linear \HNL contribution combines the mass-splitting part in \eqref{eq:0nubb HNL LO} with the interference part in \eqref{eq:0nubb HNL interference}.



\end{document}